\magnification=1200
\baselineskip=15pt
\overfullrule=0pt

\def\t{\tau}
\def\k1{k^{(1)}}
\def\k2{k^{(2)}}
\def\S{\Sigma}
\def\F{{\cal F}}
\def\M{{\cal M}}
\def\N{{\cal N}}
\def\G{\Gamma}
\def\l{\lambda}
\def\p{\partial}
\def\a{\alpha}
\def\b{\beta}
\def\r{{\rm Res}}
\def\2{{1\over 2}}
\def\3{{3\over 2}}
\def\4{{1 \over 4}}
\def\L{{\rm log}\,}
\def\Re{{\rm Re}}
\def\Im{{\rm Im}}

\rightline{UCLA/97/TEP/18}
\rightline{Columbia/Math/97}
\bigskip

\centerline{{\bf CALOGERO-MOSER SYSTEMS IN SU(N) SEIBERG-WITTEN THEORY}
\footnote*{Research supported in part by
the National Science Foundation under grants PHY-95-31023 and DMS-95-05399.}}

\bigskip
\bigskip
\centerline{{\bf Eric D'Hoker} ${}^1$ 
            {\bf and D.H. Phong} ${}^2$}
\bigskip
\centerline{${}^1$ Department of Physics}
\centerline{University of California, Los Angeles, CA 90024, USA}
\bigskip
\centerline{${}^2$ Department of Mathematics}
\centerline{Columbia University, New York, NY 10027, USA}

\bigskip
\bigskip
\bigskip

\centerline{\bf ABSTRACT}

\bigskip

The Seiberg-Witten curve and differential for ${\cal N}=2$ supersymmetric $SU(N)$ 
gauge theory, with a massive hypermultiplet in the adjoint representation of the
gauge group, are analyzed in terms of the elliptic Calogero-Moser integrable
system. A new parametrization for the Calogero-Moser spectral curves is
found, which exhibits the classical vacuum expectation values of the scalar
field of the gauge multiplet. The one-loop perturbative correction to the
effective prepotential is evaluated explicitly, and found to agree with
quantum field theory predictions. A renormalization group equation for the
variation with respect to the coupling is derived for the effective
prepotential, and may be evaluated in a weak coupling series using residue
methods only. This gives a simple and efficient algorithm for the
instanton corrections to the effective prepotential to any order.
The 1- and 2- instanton corrections are derived explicitly. Finally, it is
shown that certain decoupling limits yield ${\cal N}=2$ supersymmetric theories
for simple gauge groups $SU(N_1)$ with hypermultiplets in the fundamental
representation, while others yield theories for product gauge groups
$SU(N_1) \times \cdots\times SU(N_p)$, with hypermultiplets in fundamental
and bi-fundamental representations. The spectral curves obtained this way for
these models agree with the ones proposed by Witten using D-branes and
M-theory.

\vfill\break

\centerline{\bf I. INTRODUCTION}

\bigskip

The low energy effective theory for $\N=2$ supersymmetric gauge
theories, with a hypermultiplet in the adjoint representation of the gauge
group, was constructed for $SU(2)$ gauge group in the original paper by
Seiberg and Witten [1]. It was subsequently constructed for
arbitrary $SU(N)$ gauge groups by Donagi and Witten [2], who
also showed on general grounds that $\N=2$ supersymmetric gauge 
theories are described by integrable Hamiltonian systems. In particular, the 
integrable system corresponding to the $SU(N)$ theory
with matter in the adjoint representation
was identified as an $SU(N)$ Hitchin system, that is, a completely
integrable model arising from a two-dimensional $SU(N)$ gauge theory.
A possible close relation between the spectral curves
proposed by Donagi and Witten and the spectral curves of elliptic 
Calogero-Moser systems [3]
was suggested by Martinec in [4], and established
explicitly by Krichever in unpublished work.

\medskip

The goal of the present paper is to analyze
carefully the $\N=2$ supersymmetric $SU(N)$ gauge theory
with a hypermultiplet in the adjoint representation, 
using the elliptic Calogero-Moser integrable system. This
is the dynamical system
$$
p_i=\dot x_i,\qquad \dot p_i
= m^2 \sum_{j\not=i}\wp '(x_i-x_j),\quad 1\leq i,j\leq N.\eqno(1.1)
$$
where $\wp(z)$ is the Weierstrass $\wp$-function. It admits the Lax representation $\dot L=[M,L]$, with
$L(z)$, $M(z)$ given by $N\times N$ matrices
depending both on the dynamical variables
$x_i,p_i$ and a spectral parameter $z$ living on a torus
$\S$ (c.f. (2.11) below). The complex 
modulus $\tau$ of the torus $\Sigma$ is related to the
gauge coupling $e$ and the $\theta$-angle of the gauge theory by
$$
\tau = {\theta \over 2 \pi} + { 4 \pi i \over e^2}.
\eqno(1.2)
$$ 
The spectral curve of the elliptic Calogero-Moser system is given by
$$
R(k,z)\equiv \det (k I - L(z)) =0
\eqno(1.3)
$$
and provides the curve of the $\N=2$ theory, while the Seiberg-Witten
differential is naturally constructed out of the spectral parameter $z$ by
$d\lambda = k dz$. This construction fits naturally in the general framework
for integrable systems and Seiberg-Witten differentials proposed in [5][6]
(c.f. Section II).

\medskip

The case of the adjoint representation presents however
a number of distinctive features which set it apart from
the case of the fundamental representation, for which
a well-developed machinery is now available [7,8,9] (see also
[10], where an extensive list of references can be found).
Indeed, the spectral curves in the fundamental representation
case can be motivated by a weak-coupling analysis,
in which the classical order parameters
of the four-dimensional gauge theory 
(i.e. the vacuum expectation values of the scalar field
belonging to the gauge multiplet) are apparent.
In the adjoint representation case, there is at the present
time no such argument. Rather, the match between the four-dimensional
gauge theory and the two-dimensional integrable model
was found via indirect arguments, where
the order parameters are difficult to recognize even
in the weak-coupling limit. 

\medskip 

From the viewpoint of Calogero-Moser
systems, the $N$-dimensional family of spectral curves (1.3) is parametrized
by an overdetermined set of $2N$ phase variables $x_i,p_i$,
and it is difficult to identify in terms
of the $x_i,p_i$ the crucial $N$ variables which correspond to
the order parameters of the gauge theory.
Thus a key ingredient of our approach is a convenient
parametrization of the Calogero-Moser spectral curves
by a single monic polynomial $H(k)=\prod_{i=1}^N(k-k_i)$ of degree $N$,
whose zeroes $k_i$ are essentially the classical 
order parameters of the gauge theory. More precisely,
if we set
$$
f(k,z)\equiv R(k-m\p_z\L\vartheta_1({z\over 2\omega_1}|\tau),z),
$$     
then the elliptic Calogero-Moser spectral
curves are characterized by
$$
f(k,z)={1\over\vartheta_1({z\over 2\omega_1}|\tau)}
\vartheta_1\big({1\over 2\omega_1}
\{z-m{\partial \over \partial k}\} \big | \tau\big)H(k),
\eqno(1.4)
$$
and the classical order parameters are given by
$$
lim_{q\rightarrow0}{1\over 2\pi i}\oint_{A_i}d\l
=k_i-\2 m,
\eqno(1.5)
$$
where $m$ is the mass of the hypermultiplet.
This is established in Section III.

\medskip

With the parametrization (1.4),
it is then possible to set up
a systematic expansion at weak
gauge coupling of the quantum order parameters and their duals, 
and thus of the effective prepotential. Using 
the methods developed in [7], we derive in this way the
perturbative contribution $\F^{\rm pert}$ to the effective prepotential
$$
\F ^{{\rm pert}} =  -{1 \over 8 \pi i} \sum _{i,j =1} ^N \{
(a_i - a_j )^2 \log (a_i - a_j )^2 -(a_i - a_j +m )^2 
\log (a_i - a_j+m )^2\}
\eqno(1.6)
$$
It agrees with the well-known formula
obtained from direct 
quantum field theory calculations, and confirms
that the integrable model corresponding to the
$\N=2$ supersymmetric gauge theory with matter
in the adjoint representation is indeed the elliptic
Calogero-Moser system. We also evaluate
the 1-instanton correction, whose 
calculation has not yet been performed by
direct instanton methods (Section IV).
\medskip
In Section V, we derive a renormalization group (RG) type equation for the
variation of the prepotential with respect to the complex gauge coupling
$\tau$ :
$$
{\p \F \over \p \tau} = {1 \over 4 \pi i} \sum _{j=1} ^N \oint  _{A_j} k^2 dz
\eqno(1.7)
$$
Remarkably, the right hand side coincides with the
Hamiltonian of the Calogero-Moser system. Such a relation
also exists for theories with matter in the fundamental
representation [11][6]. It suggests a more direct
link between $\N=2$ supersymmetric gauge theories and integrable
models than just coincidence of spectral curves: {\it the RG beta
function of the four-dimensional theory
is given by the Hamiltonian of the two-dimensional
integrable model.} An important feature of (1.7) is that 
the beta function may 
be evaluated in a series expansion at weak
coupling, using residue
methods only. As a consequence, instanton corrections
can be routinely calculated to an arbitrarily high order.
We illustrate the process by deriving both
the 1-instanton and the 2-instanton corrections to the
prepotential. As expected, the 1-instanton answer agrees
with the formula obtained earlier in Section IV.

\medskip

Finally, in Section VI, we take various decoupling limits in which the
hypermultiplet mass parameter $m$ is taken to infinity. The simplest limit 
is when the full hypermultiplet is decoupled, yielding the pure $\N=2$ 
supersymmetric $SU(N)$ gauge theory. We verify that the curve as well as the
effective prepotential converge towards the forms for the pure theory. 
However, a number of more subtle limits may be taken, in which some of the
order parameters are taken to infinity as well. In general, this will yield
an $\N=2$ supersymmetric theory whose gauge group is a product $SU(N_1)
\times \cdots
\times SU(N_p)$ with hypermultiplets in fundamental and bi-fundamental
representations of the gauge group.  These theories appear to be special 
cases of the general models with product $SU(N)$ groups which were recently
solved by Witten using M-theory and D-brane technology [12] (see
also the recent work of Katz, Mayr, and Vafa [25] based rather
on compactifications of Type II strings on Calabi-Yau manifolds). Our models
appear with specific relations between the $N_i$, as well as
between their gauge couplings. In particular, we recover in this
way theories with gauge group
$SU(N_c)$ and with $N_f$ hypermultiplets in the fundamental representation.
We note that, conversely, Witten has shown in [12] how to recover
the curves of [2] from M-theory.

\vfill\break

\centerline{\bf II. THE SPECTRAL CURVES FOR THE ADJOINT REPRESENTATION}

\bigskip

The main goal of this section is to provide
a survey of the several equivalent descriptions of the
candidates for the spectral curves of the $\N=2$
supersymmetric gauge theory with a hypermultiplet
in the adjoint representation. 

\medskip

We begin by reviewing the 
original description and motivation of Donagi and Witten [2]. The idea 
in their work is to
view the theory as an $\N=4$
theory with a bare mass term breaking the $\N=4$ supersymmetry to an $\N=2$
supersymmetry, and to deform correspondingly the spectral curves for the
$\N=4$ theory. In the Coulomb phase of $\N=2$ supersymmetric $SU(N)$ gauge
theories, the moduli space of vacua is parametrized by 
the vacuum configurations of
the scalar field $\phi$ in the $\N=2$ gauge multiplet.
At the vacuum, $\phi$ is constant and lies in the Cartan subalgebra,
$\phi=\sum_{i=1}^{N-1}a_iH_i$.
In the $\N=4$ supersymmetric case,
the metric on the space of vacua $\{a_i\}$
receives no quantum corrections and is given by
$$
d s^2=({\rm Im}\,\tau)\sum_{i=1}^Nda_i\overline{da_i}
\eqno(2.1)
$$
where $\tau$ is the microscopic gauge coupling.
A key observation of [2] is that this metric also arises by a Hitchin
construction, i.e., an integrable model
based on a $two-dimensional$
gauge theory [13]. Let $\S$ be the torus with modulus $\tau$, 
and define the space
$X$ by 
$$
X=\{(A,\Phi);\overline{\nabla}_A\varphi=0\}/H,
\eqno(2.2)
$$ 
where $A$ is a connection on a flat $SU(N)$
bundle over $\S$, $\varphi$ is a (1,0) form,
both with values in the adjoint representation, and $H$ is the group
of (complexified) gauge transformations. Since the space of connections 
modulo $H$ is the space $\M$ of holomorphic $SL(n,{\bf C})$ bundles over $\S$,
and since the cotangent space of $\M$ is the space of holomorphic one-forms
valued in the adjoint representation, $X$ can be recognized as $T^*(\M)$. 
As such, it admits a symplectic form $\omega$. 
More concretely, the fundamental group of $\S$ is abelian,
and $\M$ is the same as the moduli space of holomorphic $ U(1)^{N-1}$
bundles over $\S$, divided by the Weyl group $W$, 
$\M=\S^{N-1}/W$. 
At generic points of $\M$, it follows from the condition
$\overline{\nabla}_A\varphi=0$ that $\varphi$ reduces to a constant function
with values in the Cartan subalgebra ${\bf C}^{N-1}$.
Thus we can write
$$
X=(\S^{N-1}\times{\bf C}^{N-1})/W\eqno(2.3)
$$
If we represent each copy of $\S$ in (2.3)
by points $(x_k,y_k)$ satisfying the equation 
$$
y^2=\prod_{i=1}^3(x-e_i),
\eqno(2.4)
$$
the symplectic form $\omega$ can be expressed as
$$
\omega=\sum_{i=1}^{N-1}{dx_i\over y_i}\wedge da_i
\eqno(2.5)
$$
Now, locally, we can view $X$ as a fibration over the base ${\bf C}^{N-1}$,
with the Abelian variety $\S^{N-1}$ as fiber. Let the polarization
of $\S^{N-1}$ be denoted by $t$ ($t$ is a two-form
which corresponds, physically, to the
pairing between the electric and magnetic charges of the $N-1$ $U(1)$
factors). Then the vacua metric (2.1) can be obtained by integrating
the $N\times N$ form $t^{N-2}\wedge\omega$ along the fiber.

\medskip

As the hypermultiplet acquires a non-vanishing mass $m$, the consistency of
the mass formula for BPS states requires
that (the cohomology class of) $\omega$ vary linearly with $m$.
This is implemented by deforming the above picture (2.2) to
configurations $\varphi$ on $\S$ admitting a ``pole"
with residue $m\mu$ at a point $P$ in $\S$
$$
\overline{\nabla}_A\varphi(z)=m\mu\,\delta(z,P)
\eqno(2.6)
$$ 
Here $\mu$ is a matrix in the adjoint
representation. (Strictly speaking, as we shall see below, this relation can 
only be implemented after a {\it singular} gauge transformation on $\varphi$.
This is as it should be, since $\varphi$ has a single singularity,
and hence must have an essential singularity there.) The condition
that this new space still have the same dimension as $X$ forces
$\mu$ to be a diagonalizable matrix with eigenvalues $1,\cdots,1,-(N-1)$.
The family of spectral curves $\G$ for the theory is then given by
the Riemann surfaces of equation
$$
R(k,z)\equiv\det (kI-\varphi(z))=0.
\eqno(2.7)
$$
where $I$ is the $N\times N$ identity matrix.   
Donagi and Witten provide an algorithm for 
writing the equation for $\G$ under the
form
$$
P_N+A_2P_{N-2}+\cdots+A_NP_0=0
\eqno(2.8)
$$
where the $P_i=P_i(x,y,k)$ are specific polynomials in $x,y,k$, monic
of degree $i$ in $k$, with $\tau$-dependent coefficients. The $N-1$
coefficients $A_i$ are identified with the order parameters of the
four-dimensional gauge theory [2].

\medskip

In the formalism of [5][6], the moduli space of vacua of an $\N=2$
supersymmetric
$SU(N)$ gauge theory is to be represented as a leaf in the space
$\M_g(n_{\a} ,m_{\a})$ of triples $(\G,E,Q)$, 
where $\G$ is a Riemann surface
of genus $g$, and $E$ and $Q$ are Abelian integrals with poles
of order at most $n_{\a}$ and $m_{\a}$ respectively at a fixed
number $N$ of punctures $P_{\a}$, $1\leq \a\leq N$. 
Furthermore, the Seiberg-Witten form $d\l$ is
represented by the meromorphic one-form $d\l=QdE$.  
In view of the
preceding description of $\G$, we take the number $N$ of punctures
as well as the genus $g$
to be the $N$ of $SU(N)$, $n_{\a}$ to be $-1$ (i.e., the Abelian
differential $dE$ is actually a holomorphic Abelian differential
$dz$ on $\G$), and $Q$ to be a meromorphic {\it function} $Q=k$,
with poles of order $m_{\a}=1$ at each $P_{\a}$.
The moduli space $\M_g(n_{\a},m_{\a})$ has dimension [5]
$$
dim \M_g(n_\alpha ,m_\alpha )=5g-3+3N+\sum_{\a=1}^N(n_{\a}+m_{\a})=
8N-3
$$
Let $2\omega _1,2\omega_2$ be the periods of the torus $\S$, 
(with modulus $\tau=\omega_2/\omega_1$) 
considered as fixed parameters, and let
${\cal L}$ be the lattice $2\omega _1{\bf Z}+2\omega _2{\bf Z}$. 
Then an $N$-dimensional family of Riemann surfaces
$\G$ with all the properties of (2.7) is given by the following leaf
in the canonical foliation of $\M_g(n_{\a},m_{\a})$
$$
\matrix 
{\r_{P_{\a}}dk=0,&\ 1\leq \a\leq N, & \quad \oint_{C}dk=0\cr
& & \cr
\int_{P_1}^{P_{\a}}dz\in{\cal L},&\ 2\leq\a\leq N, & \quad 
\oint_Cdz\in {\cal L}\cr
&& \cr
\r_{P_{\a}}(kdz)=-m,&\ 2\leq \a\leq N,&{\qquad \r_{P_1}(kdz)=m(N-1)}.\cr}
\eqno(2.9)
$$
We note that this accounts for $7N-3$ conditions.
The remaining $N$ parameters in $\M_g(n_\a ,m_\a )$, which are thus
the parameters for the leaf, are provided by the $A_j$ periods
$$a_j={1 \over 2 \pi i}\oint_{A_j}d\l,
$$ 
where $A_j,B_j$ are a canonical homology
basis on $\G$. We observe that there are $N$ of them,
and thus one more than the dimension of the moduli
space of vacua of the $SU(N)$ gauge theory.
This is as it should be,
since the $SU(N)$ traceless condition $\sum_{i=0}^Na_i=0$
is yet to be imposed. However,
the above set-up is clearly invariant
under shifts of the meromorphic function
$k$ by a constant, and it is this arbitrariness 
which allows us to
impose the traceless condition.  

\medskip

Remarkably, as anticipated by Martinec [4] and established
by Krichever,
the spectral curves specified by (2.7) and (2.9)
are precisely the spectral curves written down
much earlier by Krichever [3] for the elliptic
Calogero-Moser system
$$
p_i=\dot x_i,\qquad \dot p_i
= m^2 \sum_{j\not=i}\wp '(x_i-x_j),\quad 1\leq i,j\leq N,
\eqno(2.10)
$$
where $\wp(z)$ is the Weierstrass $\wp$-function
with periods $2\omega_1, 2\omega_2$. In fact, it
was shown by Krichever in [3] that (2.10) is equivalent
to the Lax equation $\dot L=[M,L]$, where
$(L(z),M(z))$ is the following Lax pair with spectral parameter $z\in\S$
$$
\eqalign{
L_{ij}(z)
&=p_i\delta_{ij}-m(1-\delta_{ij})\Phi(x_i-x_j,z)\cr
M_{ij}(z)
&=m\delta_{ij}\sum_{k\not=i}\wp(x_i-x_k)
+m(1-\delta_{ij})\Phi'(x_i-x_j,z)
\cr}
\eqno(2.11)
$$
Here $\Phi(x,z)$ is the function defined by
$$
\Phi(x,z) \equiv
{\sigma(z-x)\over\sigma(z)\sigma(x)}e^{\zeta(z)x}
\eqno (2.12)
$$
where $\sigma (z) $ and $\zeta (z)$ are the Weierstrass functions, given by
$$
\sigma(z)=z\prod_{n_1,n_2}(1-{z\over n_1\omega_1+n_2\omega_2})
{\rm exp}({z\over n_1\omega_1+n_2\omega_2}+{1\over 2}
({z\over n_1\omega_1+n_2\omega_2})^2),
$$
and $\zeta(z)=\p_z{\rm log}\sigma(z)$. Recall that we have 
$\wp(z)=-\p_z\zeta(z)$ and $\sigma(z)=z+{\cal O}(z^5)$.
Also, $\Phi ' (x,z)$ denotes the $x$-derivative of $\Phi (x,z)$, and the key
identity on the function $\Phi (x,z)$ used in showing that (2.10) follows from
the Lax equation is
$$
\Phi (x,z) \Phi '(y,z) - \Phi (y,z) \Phi '(x,z) 
=(\wp (x) - \wp(y) ) \Phi (x+y,z).
$$

\medskip

The functions $\Phi(x,z)$ and $L(z)$ 
are doubly periodic in $z$ with periods
$2\omega _1$ and $2\omega_2$ and thus well defined on the torus $\S$. 
$L(z)$ has an 
essential singularity at $z=0$, of the form
$$
L_{ij}(z)\sim m(1-\delta_{ij}){e^{-{1\over z}(x_i-x_j)}\over z}
\eqno(2.13)
$$
However, under the local and singular gauge transformation 
$L\rightarrow \tilde L=G^{-1}L(z)G$, with
$G_{ij}=\delta_{ij}e^{\zeta (z)x_i}$, $L$ is transformed into
$$
\tilde L_{ij}(z)=p_i\delta_{ij}-m(1-\delta_{ij}){\sigma(z-x_i+x_j)
\over\sigma(z)\sigma(x_i-x_j)}
\eqno(2.14)
$$
which is meromorphic near $z=0$. Thus the equation of the spectral curve $\G$
for the elliptic Calogero-Moser system
$$
R(k,z)=\det (kI-L(z))=0
\eqno(2.15)
$$
is doubly periodic as well as meromorphic in $z$. The
invariance of the theory under shifts of $k$ by constants
corresponds in this set-up to the freedom
of adding a constant matrix to $L(z)$, or equivalently,
to shift all momenta $p_i$. 
The spectral curve $\G$ is a $N$-sheeted covering of the torus
$\S$. The holomorphic Abelian differential $dz$ on $\S$ pulls back
to a holomorphic Abelian differential on $\G$, which we still
denote by $dz$. The solution $k$ of (2.15) defines a single-valued
meromorphic function on $\G$. Clearly, all its poles $P_1,\cdots, P_N$
lie on top of the pole at $z=0$ of $R(k,z)$. This shows that
the second equation in (2.9) must be satisfied. It is also easy to verify
all the other constraints in (2.9). 

\bigskip
\bigskip

\centerline{\bf III. THE CLASSICAL ORDER PARAMETERS}

\bigskip

In the case of $\N=2$ supersymmetric gauge theories with classical gauge
groups, the classical order parameters of the theory are apparent
from the equation of the spectral curve. For example, for
$SU(N)$ theories with matter in the fundamental representation,
they are recognizable as the parameters $\bar a_k$ in the equation [7]
$$
y^2=\prod_{i=1}^N(x-\bar a_k)^2-\Lambda^{2N_c-N_f}\prod_{j=1}^{N_f}(x+m_j)
\eqno (3.1)
$$
The issue is more subtle in the adjoint representation case. Presumably the
coefficients $A_k$ of the equation (2.8) are gauge invariant polynomials in
the scalar field $\phi$ [2], although this is not entirely manifest.
In the Calogero-Moser formalism, we have $2N$ parameters
$x_i,p_i$ which are natural from the dynamical system but not
from the gauge theory viewpoint. We can also try to characterize the polynomial 
$R(k,z)$ by its coefficients, which are elliptic functions and hence
linear combinations of the Weierstrass $\wp$-function and its derivatives. However,
the number of these terms exceeds $N$, and linear constraints
have to be imposed on them [14], which make their gauge theoretic
interpretation obscure. Thus a first basic step of our approach
is a more appropriate parametrization for (2.8)(2.9)(2.15), which will
shed light on the correct order parameters. 

\medskip

To achieve this, we begin by reducing the singularity at $z=0$ in $R(k,z)$ to 
a simple pole by performing a shift in $k$. 
This can be done at the cost of replacing
the double periodicity of $R(k,z)$ by a more complicated transformation law.
We define the function $h_n(z)$ by
$$
h_n(z)
\equiv {1\over\vartheta_1({z\over 2\omega_1}|\tau)}
{\partial ^n \over \partial z^n}
\vartheta_1({z\over 2\omega_1}|\tau)
\eqno(3.2)
$$
where the Jacobi $\vartheta$-function is given by
$$
\vartheta_1(u|\tau)=\sum_{r\in\2+{\bf Z}} e^{i\pi r^2\tau+2i\pi r(u+\2)}
\eqno (3.3a)
$$
In view of the transformation laws of the $\vartheta$-function under shifts of
$u$
$$
\eqalignno{
\vartheta_1(u+1|\tau)&=e^{i\pi}\vartheta_1(u|\tau)\cr
\vartheta_1(u+\tau|\tau)&=e^{-\pi i\tau-2\pi i (u-\2)}\vartheta_1(u|\tau),
&(3.3b)\cr}
$$
we see that the functions $h_n(z)$ transform as follows
$$
\eqalignno{
h_n(z+2\omega_1)&=h_n(z)\cr
h_n(z+2\omega_2)&=\sum_{p=0}^n\pmatrix{n\cr p\cr}\beta^{n-p}h_p(z)
\qquad \qquad \beta = - {i \pi \over \omega _1}
&(3.4)\cr}
$$
In particular, we have $h_1(z+2 \omega _2) = h_1(z) + \beta$, since $h_0=1$.

\medskip

We now introduce the function $f(k,z)$ by
$$
f(k,z)\equiv R(k-mh_1(z),z)
\eqno(3.5)
$$
While $R(k,z)$ has a pole of order $N$ in $z$ at $z=0$ and is doubly
periodic in $z$, the function $f(k,z)$ is constructed to have only a simple
pole at $z=0$, at the cost of the following non-trivial transformation laws
under shifts of $z$ 
$$
\eqalignno{
f(k,z+2\omega_1)&=f(k,z)\cr
f(k,z+2\omega_2)&=f(k-\beta m,z).
&(3.6)\cr}
$$
We claim that $f(k,z)$ can be simply expressed in terms of a single 
polynomial of degree $N$ whose zeroes are linearly related to the classical
order parameters of the $\N=2$ supersymmetric gauge theory in the Coulomb
phase.  This can be seen as follows. The residue at the pole $z=0$ is a
polynomial in $k$ of degree $N$, so that the residues at the poles $z=2
\omega _1 {\bf Z} + 2  \omega _2 {\bf Z}$  are shifts of the polynomial at
$z=0$, according to the monodromy law (3.6). Thus, the pole structure of
$f(k,z)$ is completely determined by a polynomial in $k$ of degree $N$. The
relation with the classical order parameters will be established below.

\medskip

To derive a concrete relation, we make use of (3.6) to derive a
transformation law for the coefficients $f_n(z)$ in
$f(k,z)=\sum_{n=0}^Nf_n(z)k^{N-n}$ 
$$
\eqalign{
f_n(z+2\omega_1)&=f_n(z)\cr
f_n(z+2\omega_2)&=\sum_{p=0}^nf_p(z)\pmatrix{N-p\cr n-p\cr}(-\beta m)^{n-p}
\cr}
\eqno (3.7)
$$
We notice that the transformation laws in (3.7) and (3.4) are closely
related. One can show that the functions $f_n(z)$ may be expressed
as linear combinations of the functions $h_p(z)$ with $0\leq p \leq n$.
To do so, one constructs a linear combination $\tilde f _n (z)$ (with
$\beta$-dependent coefficients) of $h_p(z)$ with $0\leq p \leq n$ that has
the same transformation laws (3.7) and the same simple pole at $z=0$ as
$f_n(z)$. The difference $f_n (z) - \tilde f _n (z)$ is then holomorphic
and doubly periodic and thus must be constant. The constant may be
absorbed in the $h_0(z)=1$ term in $\tilde f_n(z)$.  

\medskip
 
Thus, $f(k,z)$ admits a decomposition under the form
$$
f(k,z)=\sum_{n=0}^Nh_n(z)Q_{N-n}(k)
\eqno(3.8)
$$
with $Q_p(k)$ a polynomial in $k$ of degree $p$. In view of (3.6) the 
$Q_p(k)$ satisfy the following recursion relation
$$
Q_p(k-\beta m)=\sum_{n=0}^p\pmatrix{N-n\cr p-n\cr}\beta^{p-n}Q_n(k).
\eqno(3.9)
$$   
In terms of the generating function $H(t,k)=\sum_{p=0}^N t^{N-p}Q_p(k)$, (3.9) is
equivalent to $H(t + \beta,k+\beta m)=H(t,k)$. Since $H(t,k)$ is
polynomial both in $t$ and in $k$, this condition requires that $H(t,k) =
H(0,k-tm)$ depends only upon the combination $k- tm$. Defining the polynomial
of a single variable $H(k) \equiv H(0, k) $, we have 
$$
H(t,k)=H(k- tm)=\sum_{p=0}^N{(-)^p(tm)^p\over p!}H^{(p)}(k),
\eqno(3.10)
$$
which identifies the polynomials $Q_n(k)$ of (3.7) as
$$
Q_{N-n}(k)={(-m)^n\over n!}H^{(n)}(k) 
\eqno(3.11)
$$
 
\medskip

In summary, we have shown that the function $f(k,z)$ can be re-expressed
in terms of a polynomial $H(k)$ of degree $N$. The coefficient of the leading
monomial is given by the general form of $f(k,z)$ and equals 1. The remaining
$N$ coefficients, or equivalently the $N$ roots of the polynomial $H(k)$,
represent the classical order parameters of the $\N=2$ supersymmetric gauge
theory. In the next section, we shall demonstrate this
fact explicitly by calculating the classical limit (${\rm Im}\, \tau \to
\infty$)  of the quantum order parameters. 

\medskip

Combining (3.2), (3.8) and (3.11), we find a simple form for
$f(k,z)$ in terms of $H(k)$ :
$$
f(k,z)={1\over\vartheta_1({z\over 2\omega_1}|\tau)}
\sum_{n=0}^N{1\over n!} {\partial ^n \over \partial z ^n}
\vartheta_1 \big ({z\over 2\omega_1}|\tau \big) \big (-m{\p \over
\p k}\big )^nH(k)
\eqno(3.12)
$$
or, even more succinctly,
$$
f(k,z)={1\over\vartheta_1({z\over 2\omega_1}|\tau)}
\vartheta_1\big({1\over 2\omega_1}
\{z-m{\partial \over \partial k}\} \big | \tau\big)H(k).
\eqno(3.13)
$$

\medskip

The original equation for the curve $\Gamma$ was given by $R( \tilde k,z)=0$,
and the associated Seiberg-Witten differential is then $d\lambda = \tilde k 
dz$. It is convenient to translate these expressions in terms of variables
that are naturally adapted to the function $f(k,z)$ instead. This is
achieved by setting $k\equiv \tilde k + m h_1 (z) + \2 \beta m$, so that the
equation for $f$ and the Seiberg-Witten differential in terms of $k$ become 
$$
\eqalign{
0 = & f(k- \2 \beta m, z)  \cr
d \lambda =  \tilde k dz = & kdz - m h_1 (z) dz - \2 \beta m dz \cr}
\eqno (3.14)
$$
It is often convenient to ignore the term $-\2 \beta m dz$ in $d\lambda$,
as we shall do in the beginning of Section IV.(c).
This term contributes to $a_i$ and $a_{Di}$ a constant $i$-independent shift,
whose effect on the prepotential is easily read off (c.f. (4.37) and (4.38)
below). The simple poles at the $N$ lifts of the pole at $z=0$ are clearly
exhibited by this expression in the part $-mh_1(z)dz$, with a remaining pole
on just a single sheet left in the part $kdz$. The residue  at this
remaining pole is readily checked to be $Nm$, in agreement with the residue
conditions of (2.9). (Notice that we could have chosen the period
$2\omega _1 = -2 \pi i$, so that $\beta =1$; we shall instead keep track of
general $\beta$ throughout and show as a check that physical quantities, such as
the prepotential, are independent of $\beta$, when expressed in terms of the
quantum order parameters.)

\medskip

Using the series expansion for $\vartheta$-functions, given by (3.3a), we
find that the equation for the spectral curve $f( k -\2 \beta m,z)=0$ admits
the following series expansion in powers of $q=e^{2i\pi\tau}$
$$
\sum_{n\in{\bf Z}}(-)^nq^{\2 n(n-1)}
e^{\beta nz}H( k -\beta mn)
=0.
\eqno(3.15)
$$
This series is remarkably sparse : the orders to which corrections
occur grow quadratically~!

\bigskip
\bigskip

\centerline{\bf IV. THE QUANTUM ORDER PARAMETERS AND}
\centerline{\bf THE EFFECTIVE PREPOTENTIAL}

\bigskip

We are now ready to evaluate the quantum order parameters $a_i$, their duals
$a_{Di}$ and the prepotential ${\cal F}(a _i ,\tau)$, given by
$$
a_i={1\over 2\pi i}\oint_{A_i}d\lambda,\qquad
a_{Di}={1\over 2\pi i}\oint_{B_i}d\lambda,\qquad
a_{Di}={\p{\cal F}\over \p a_i},
\eqno(4.1)
$$
in the weak coupling limit. Recall that the coupling $\tau$ is related to the
gauge coupling $e$ and $\theta$-angle of the microscopic gauge theory by
$$
q=e^{2i\pi\tau}=e^{-{8\pi^2\over e^2}+i\theta}.
$$
Thus the weak-coupling limit corresponds to  vanishing $q$.
Our main goals are
\medskip 
\item{(a)} to justify the zeroes of the polynomial $H(k)$ introduced in the
previous section as classical order parameters;
\medskip 
\item{(b)} to establish that the spectral curves for the elliptic
Calogero-Moser system are the correct Seiberg-Witten spectral curves for the
$\N=2$ supersymmetric $SU(N)$ gauge theory with  matter in the adjoint
representation. In particular, the prepotential ${\cal F}$ of (4.1) must
exhibit the correct logarithmic singularities which would arise from the
one-loop perturbative effects of the supersymmetric gauge theory;
\medskip
\item{(c)} to evaluate the contributions to ${\cal F}$
of instanton processes. In this section, we shall carry this out
directly from (4.1) to one instanton order. In the next, we shall derive a
renormalization group equation for ${\cal F}$ from which the one-
and two-instanton contributions can be read off at once. The two independent
methods also serve as mutual checks for each other.

\medskip

Before giving the details of the calculation, it may be helpful
to discuss some aspects of our method
and of the underlying geometry.

\medskip

$\bullet$ The Riemann surface $\G$ defined by the equation (2.9) is an
$N$-sheeted covering of the torus $\S$. If we represent $\S$ by the
lattice $2\omega_1{\bf Z}+2\omega_2{\bf Z}$ and let $A$ and $B$ be the usual canonical homology
cycles on $\S$, then a homology basis $(A_i,B_i)$ for $\G$ is just given by 
the lifts
$A_i,B_i$ to each sheet of the $A,B$ cycle on the base $\S$.

\medskip

$\bullet$ At the classical limit $q=0$, the cycle
$A$ shrinks to a point, and the base $\S$ degenerates into a sphere
with two marked punctures (to be identified with the shrunken
cycle $A$). If we set 
$$w=e^{\b z},
\eqno(4.2)
$$
the two marked punctures
are given by $w=0$ and $w=\infty$. Upstairs, the cycles $A_i$ also 
degenerate into $2N$ punctures, with a pair on each sheet. If we also view 
each sheet as
a sphere with two marked punctures, then the shrunken $A_i$ corresponds
to two punctures $k_i$ and $k_i+\beta m$, which lie respectively over
the punctures $w=0$ and $w=\infty$ on the base. We shall
show that the punctures $k_i$ are precisely the zeroes of the 
polynomial $H(k)$
$$
H(k)=\prod_{i=1}^N(k-k_i)\equiv (k-k_i)H_i(k)
\eqno(4.3)
$$
In (4.3), we have taken the opportunity to define the polynomials $H_i(k)$,
which we shall use shortly.

\medskip

$\bullet$ As $q$ moves away from 0, each of the punctures 
in both the base $\S$ 
and the covering $\G$ opens
into a cut, the edges of which reconstitute the $A$ cycle downstairs
and the $A_i$ cycles upstairs upon regluing.
Let the cut near $k_i$ run from a point $k_i^-$ to another point $k_i^+$.
The points $k_i^{\pm}$ are identified as turning points, i.e.,
solutions of the equation
$$
{dk\over dz}=0
\eqno(4.4)
$$   
which we can solve perturbatively in $q$. They lie respectively over the end
points $w^{\pm}$ of the cut downstairs in $\S$. Similarly, we have cuts at
the other end going from $k_i^-+\b m$ to $k_i^++\b m$. The $B_i$ cycle in
$\G$ can be represented by a path on the $i$-th sheet going from $k_i^+ +\b
m$ to $k_i^+$. This is the path we shall use for evaluating $a_{Di}$. For
$a_i$, we shall use any simple closed path surrounding the cut near
$k_i$ in the clockwise direction. With these conventions, the $A_i$ and 
$B_i$ cycles have canonical intersection form $\# (A_i, B_j)= \delta _{ij}$.
The evaluation of $a_i$ is much simpler than 
that of $a_{Di}$, since perturbatively in $q$,
the cut shrinks to a point, and the residue formula applies.

\medskip

$\bullet$ Our calculations will be carried out on each sheet upstairs
rather than downstairs, i.e., in terms of the variable $k$ rather
than the variable $z$. This is suggested by the role of $k_i$
as classical order parameters, and is possible since the cycles
$A_i$ and $B_i$ both lie on a single copy of the complex plane.
For this however, we need a careful expansion of the
differential $dz$ in terms of $dk$, which is derived in (4.13)(4.16)
and (4.19) below.
Our starting point is the equation $f(k,z)=0$ of (3.15), which we shall
re-express in terms of the variable $w$ : 
$$
\eqalign{
H(k)-wH(k-\b m)
+\sum_{n=1}^{\infty}
& (-)^nq^{\2n(n+1)} \big [w^{-n}H(k+\b m n) \cr
&-w^{n+1}H(k-\b m-\b m n) \big ]=0.\cr}
\eqno (4.5)
$$
We note that at the degeneration point $q=0$, the zeroes $k_i$ of $H(k)$
and $k_i+\b m$ of $H(k-\b m)$ correspond indeed respectively to the nodes 
$e^{\b z}=w=0$ and $e^{\b z}=w=\infty$ of the degenerating torus $\S$.

\medskip

We now carry out the above program, up to one-instanton order,
that is, up to and including ${\cal O}(q)$ terms. 

\vfill\break

\noindent{\bf (a) The Turning Points $k_i^{\pm}$}

\medskip

The turning points $k_i^{\pm}$ are obtained by imposing the constraint
$dk/dw=0$ together with the equation $f(k - \2 \beta m,z)=0$ of the curve. 
Up to and including first order in $q$, they are then solutions of the system
$$
\eqalignno{
H(k)-wH(k-\b m)-{q\over w}H(k+\b m)+qw^2H(k-2\b m)&=0\cr
-H(k-\b m)+{q\over w^2}H(k+\b m)+2qwH(k-2\b m)
&=0&(4.6)
\cr}
$$
Solving for $H(k)$ and $H(k-\b m)$, this system can be put under the form
$$
\eqalignno{
H(k-\b m)&={q\over w^2}H(k+\b m)
+2qwH(k-2\b m)\cr
H(k)&=2{q\over w}H(k+\b m)+qw^2H(k-2\b m)
&(4.7)\cr}
$$
Near the solution $w=0$ for $q=0$, we note that $w \sim q^\2$ so that the 
terms suppressed by a power of $w^3$ can be ignored. Eliminating $w$ between
the remaining terms yields an equation for $k$ 
$$
H(k) ^2 = 4q H(k+\beta m) H(k-\beta m)
\eqno (4.8)
$$
With the help of the function $H_i(k)$, defined in (4.3), we rewrite (4.8)
as an iterative equation for $k$ :
$$
(k-k_i)^2 = 4 q {H(k+\beta m ) H(k-\beta m) \over H_i (k) ^2}
\eqno(4.9)
$$
This equation admits two solutions, $k_i ^\pm$, which correspond to the
two end points of the branch cut associated with the $A_i$-cycle on $\Gamma$. We
shall need their explicit forms up to order ${\cal O}(q)$, and also present 
the associated value of $w$, to be made use of later on :
$$
\eqalign{
k_i^{\pm} = & k_i\pm q^{\2}k_i^{(1)}+qk_i^{(2)},\cr
w(k_i ^\pm) = & 2q H(k_i ^\pm +\b m) H(k_i ^\pm ) ^{-1}.\cr}
\eqno(4.10)
$$
Here, we have introduced the following $q$-independent functions
$$
\eqalignno{
k_i^{(1)}&=2{H^{\2}(k_i-\b m)H^{\2}(k_i+\b m)\over
H_i(k_i)}\cr
k_i^{(2)}&=2{d\over dk}\bigg(
{H(k-\b m)H(k+\b m)\over H_i(k)^2}\bigg) \bigg |_{k=k_i}
&(4.11)
\cr}
$$
Near the solution  $w=\infty$ of the degenerate case $q=0$,
we note that $w\sim q^{-\2}$, so that the terms suppressed by a power of 
$w^{-3}$ may be ignored in (4.6). 
Proceeding as before, we find that the turning points 
occur precisely at $k_i ^\pm +\beta m$ with
$$
w(k_i ^\pm + \b m) =  {1 \over 2q} H(k_i ^\pm) H(k_i ^\pm -\b m)^{-1} = 
{1\over q}w(k_i ^\pm)
\eqno(4.12)
$$
where the last equality is as expected from the definition of $w=e^{\beta z}$.

\bigskip

\noindent{\bf (b) Expansion of the Differential $dz$}

\medskip

Our next step is to rewite the variable $z$ (equivalently $w=e^{\b z}$) 
in terms of the variable $k$ on each sheet. 
It is convenient to introduce the new variable $y$ by
$$
w=y{H(k)\over H(k-\b m)}.
\eqno(4.13)
$$ 
In terms of $y$, the equation (4.5) can be rewritten as a fixed point 
equation
$$
y=1+qF(y)
\eqno(4.14)
$$
where the function $F(y)$ is defined by
$$
\eqalignno{
F(y)=\sum_{n=1}^{\infty}&q^{\2n(n+1)-1}(-)^n \big[y^{-n}
\eta_n(k,\b)-y^{n+1}\eta_n(k-\b m,-\b)\big]\cr
&\eta_n(k,\b)
={H(k+\b m n)H(k-\b m) ^n\over H(k) ^{n+1}}.
&(4.15)\cr}
$$
Formally, an iterative solution to all orders is given by (c.f. [7])
$$
y=1+\sum_{n=1}^{\infty}q^ny_n, \qquad \quad
y_n={1\over n!}{\p ^{n-1}\over\p y ^{n-1}}F^n(y) \bigg | _{y=1}.
\eqno(4.16)
$$
Keeping only the expansion terms up to order $q$ will yield
$$ 
y=1-q\eta_1(k,\b)+q\eta_1(k-\b m,-\b)
$$
This approximation is valid as long as neither $\eta _1$ grows too fast as    
$q\to 0$. In particular, it is applicable in the evaluation of the
$A_i$ periods, since the integration contours can be chosen
to be at a finite fixed distance from the cuts and the points $k_i$ in 
this case. 

\medskip

However, in the evaluation of the $B_i$ periods, the points on the cuts 
can come within a distance $q^{\2}$ of the nodes $k_i$ and $k_i+\b m$. This
invalidates in this case the above naive approximation, and requires a more
careful analysis. To order up to and including ${\cal O}(q)$ we can write the
equation (4.15-16) as
$$ 
y=1-y^{-1}q\eta_1(k,\b)+y^2q\eta_1(k-\b m,-\b)
\eqno (4.17)
$$
A key observation is that $\eta_1(k,\b)=\infty$ is equivalent to
$\eta_1(k-\b m,-\beta)=0$ and vice versa. Thus the functions
$y=\2+\2\sqrt{1-4q\eta_1(k,\b)}$ and 
$y^{-1}=\2+\2\sqrt{1-4q\eta_1(k-\b m,-\b)}$
are good approximations of the solution near $k_i$ and $k_i+\b m$ 
respectively. Since they each approach 1 near the other solution,
the natural candidate for a solution valid everywhere is
$$
y
=(1+\sqrt{1-4q\eta_1(k,\b)})
(1+\sqrt{1-4q\eta_1(k-\b m,-\b)})^{-1}
\eqno (4.18)
$$
This can indeed be checked to be a solution of (4.17) up to and including
order ${\cal O}(q)$. 

We make use of the following expansion
$$
{\rm log}(\2+\2\sqrt{1-x})
=-\2\sum_{n=1}^{\infty}
{\G(n+\2)\over\G(\2)\G(n+1)}{1\over n}x^n.
$$
Rewriting $\eta_1(k,\b)$ and $\eta _1(k-\b m,-\b)= \eta _1(k-\b m, \b)$ in 
terms of $k_i^{(1)}$ and $k_i^{(2)}$, we obtain the following pole expansion
$$
(4\eta_1(k,\b))^n= \sum _{i=1} ^N \{ {1\over (k-k_i)^{2n}}(k_i^{(1)})^{2n}
+{1\over (k-k_i)^{2n-1}}2n(k_i^{(1)})^{2n-2}k_i^{(2)}+\cdots \}
$$
(where we have ignored poles of order $2n-2$ and lower). This results in
the following
expansion for ${\rm log}\,y$ to this order
$$
\eqalign{
{\rm log}\,y
=-\2\sum_{n=1}^{\infty}
{\G(n+\2)\over\G(\2)\G(n+1)}&{1\over n}q^n
\sum_{i=1}^N(k_i^{(1)})^{2n}\big[{1\over (k-k_i)^{2n}} \cr
& +2n{k_i^{(2)}\over (k_i^{(1)})^2}{1\over (k-k_i)^{2n-1}} + \cdots \ \  
-\{ k \to k-\b m \} \big ]
\cr}
\eqno (4.19)
$$
We note that, as in the case of matter in
the fundamental representation treated in [7],
we need to keep in this expansion all powers $q^n$ of $q$.
This is because the factors $(k-k_i)^{-2n}$
and $(k-k_i)^{-2n+1}$ can be as large as
$q^{-n}$ and $q^{-n+\2}$.
Together with (4.13), this provides us with the desired
expansion for $dz={1\over \b}d{\rm log}\,w$.

\bigskip

\noindent{\bf (c) Evaluation of the Periods $a_i$ and $a_{Di}$}

\medskip

Since the expansion (4.19) for the Seiberg-Witten differential
$d\l$ is an expansion in terms of poles, the methods of [7]
apply to give expansions for $a_i$ and $a_{Di}$. Following
our discussion at the beginning of this Section, we shall
reexpress the Seiberg-Witten differential $d \lambda$ of (3.14)
in terms of the variable $k$. 
As suggested after the equation (3.14), 
we can momentarily ignore the term $-\2 \beta m dz$,
and restore it only at the end, in the final
formulas for the order parameters and the
prepotential.
To lighten the notation,
we shall still use in this Section
IV.(c) the notation $d\l$, $a_i$,
and $a_{Di}$ as if no change had been made.
Only in Section IV.(d), when
an explicit distinction has to be made between
the original case and
the case where $-\2\b m$ has been dropped,
do we need to introduce a new notation
$d\tilde\l$, $\tilde a_i$, and $\tilde a_{Di}$
for the Seiberg-Witten differential and quantum
order parameters in the latter case.
We have then 
$$
\eqalignno{
d\l
&= {1 \over \b} k d \log w(k) - m h_1 (z) dz \cr
&={1\over \b}\big(
k d{\rm log}{H(k)\over H(k-\b m)}
-({\rm log}\,y)dk+d(k{\rm log}\,y)\big) - m h_1(z) dz \cr
&={1\over \b}\big(d\l^{(0)}+d\l^{(1)}+d\l^{(2)}
+d(k{\rm log}\,y)\big) - m h_1 (z) dz
&(4.20)
\cr}
$$
where we have labelled the term
$k\,d\L(H(k)/ H(k-\b m))$
by $d\l^{(0)}$, and separated the contributions of the series
(4.19) to the term $-({\rm log}\,y )dk$ into
the contribution $d\lambda^{(1)}$ for $n=1$ and the contribution
$d\l^{(2)}$ for $n\geq 2$.

\medskip

The $A_i$ periods yield the quantum order parameters $a_i$ and are readily
evaluated to order ${\cal O}(q)$
$$
a_i={1\over \b}(k_i+\2 qk_i^{(2)}).
\eqno(4.21)
$$ 
This confirms that the parameters $k_i$ of our parametrization (3.14)(4.3) 
are indeed
the classical order parameters, i.e., the limits as $q\to 0$ of the quantum
order parameters $a_i$ given by the $A_i$-periods of the Seiberg-Witten
differential. 

\medskip

We turn now to the $B_i$ periods $a_{Di}$, which we break up
in parallel with (4.20), with contributions $a_{Di}^{(p)}$ arising from the
integration of the differentials $d \lambda ^{(p)}$, $p=0,1,2$. 
$$
\eqalign{
- 2\pi i \beta a_{Di}
= &\int_{k_i^+}^{k_i^++\b m} \beta d\l \cr
=&
a_{Di}^{(0)}+a_{Di}^{(1)}+
a_{Di}^{(2)}+\int_{k_i^+}^{k_i^++\b m}\{ d(k{\rm log}\,y)-
 \beta m h _1(z)dz\} 
\cr}
\eqno(4.22)
$$
Substituting (4.19) in, all the integrals are
easily evaluated. We find
$$
\eqalignno{
a_{Di}^{(0)}
=&\sum_{j=1}^N \{ 
(k_j-\b m){\rm log}\,
{k_i^+ -k_j + \b m \over k_i^+-k_j}
+(k_j+\b m){\rm log}\,{k_i^+-k_j-\b m
\over k_i^+-k_j}\cr
& \qquad \qquad \qquad
 +\b m{\rm log}\,{k_i^+ -k_j + \b m \over k_i^+-k_j} \}\cr
a_{Di}^{(1)}
=&{1\over 4}q
\sum_{j=1}^N\bigg[ (k_j^{(1)})^2 
\big ({2\over k_i^+-k_j}-{1\over k_i^+  -k_j + \b m}
-{1\over k_i^+  -k_j - \b m}\big )\cr
&\qquad \qquad
+2k_j^{(2)}{\rm log}\,{k_i^+  -k_j + \b m \over k_i^+-k_j}
+2k_j^{(2)}{\rm log}\,{k_i^+  -k_j - \b m \over k_i^+-k_j}
\bigg ]
 & (4.23)\cr
a_{Di}^{(2)}=
&\sum_{n=2}^{\infty}
{\G(n+\2)\over\G(\2)
\G(n+1)}{1\over 2n}q^n\bigg[
{2\over 2n-1}{(k_i^{(1)})^{2n}\over (k_i^+-k_i)^{2n-1}}
+{4n\over 2n-2}{k_i^{(2)}(k_i^{(1)})^{2n-2}\over
(k_i^+-k_i)^{2n-2}}\bigg]
\cr}
$$
The $a_{Di}^{(2)}$ terms can be simplified, using the expansion
(4.10) for $k_i ^+$, and the numerical series of [7]
$$
\eqalign{
\sum_{n=2}^{\infty}{\G(n+\2)\over\G(\2)\G(n+1)}{1\over n(2n-1)}=&{3\over 2}
-2\L 2\cr
\sum_{n=2}^{\infty}{\G(n+\2)\over\G(\2)\G(n+1)}{1\over n(2n-2)}=&1-\L 2.
\cr} 
$$
One finds
$$
a_{Di}^{(2)}=
(\3-2{\rm log}\,2)q^{\2}k_i^{(1)}+ (1-{\rm log}\,2)qk_i^{(2)}
\eqno (4.24)
$$
Next, using the second expression in (4.10), we recognize the third term in
$a_{Di} ^{(0)}$ as
$$
\eqalign{
\b m\sum_{j=1}^N{\rm log}\,{k_i^+ -k_j + \b m\over k_i^+-k_j}
=& \b m \log {H(k_i ^+ + \b m) \over H(k_i ^+)} \cr
=& \b m (\log w(k_i ^+) - \log q - \log 2)\cr}
\eqno (4.25)
$$
Using (4.10) and (4.12), as well as the definition of the variable $y$ in
(4.14), we see that $y(k_i ^+) = \2$, while $y(k_i ^+ + \b m) = 2$.  With
these results, we may now easily evaluate the following integral
$$
\eqalign{
\int_{k_i^+}^{k_i^++\b m}d(k\L y)
=& (k_i^+ +\b m)\L y(k_i^+ +\b m)-k_i^+\L y(k_i^+) \cr
=& (2 k_i ^+ + \beta m ) \log 2.\cr}
\eqno(4.26)
$$
Using the definition of $h_1(z)$ in (3.2), we evaluate the integral
$$
-\beta m \int_{k_i^+}^{k_i^+ +\b m} h_1(z) dz = i \pi \beta m \tau - \beta m
\log w(k_i ^+)
\eqno (4.27)
$$
Combining these contributions, and regrouping the logarithmic terms in
$a_{Di}^{(0)}$ and $a_{Di}^{(1)}$ using (4.21), we find
$$
\eqalignno{
-2\pi i \beta a_{Di}
=&\sum_{j=1}^N\beta (a_j-m)\L({k_i^+-k_j+\b m\over k_i^+-k_j})
+\beta (a_j+m)\L({k_i^+-k_j-\b m \over k_i^+-k_j})\cr
&\quad+{1\over 4}q
\sum_{j=1}^N\big({2(k_j^{(1)})^2\over k_i^+-k_j}
-{(k_j^{(1)})^2\over k_i^+ -k_j + \b m }-
{(k_j^{(1)})^2\over k_i^+  -k_j - \b m}\big)\cr
&\quad +\3q^{\2}k_i^{(1)}+qk_i^{(2)} + 2\beta \log2 a_i - \pi i \beta m \tau
&(4.28)
\cr}
$$
Next, we evaluate (4.8) on $k=k_i ^+$. By taking the logarithm, it follows 
that
$k_i^+$ must satisfy the following identity
$$
0= \sum_{j=1}^N \{ 
 \L ({k_i^+-k_j+\b m \over k_i^+-k_j})
+\L ({k_i^+-k_j-\b m \over k_i^+-k_j}) \}
+\L q+2\L 2.
\eqno(4.29)
$$
Dividing all of equation (4.28) by a factor of $-\beta$ and adding (4.29)
multiplied by a factor of $ a_i$ to (4.28), allows us to recast (4.28) as
$$
\eqalignno{
2 \pi i  a_{Di}
=&
a_i\L q
-{3 \over 2 \beta}q^{\2}k_i^{(1)}
-{1 \over \beta} qk_i^{(2)}  + \pi i m \tau\cr
&-\sum_{j=1}^N \{ (a_j-a_i-m)\L({k_i^+-k_j+\b m\over k_i^+-k_j})
+(a_j-a_i+m)\L({k_i^+-k_j-\b m \over k_i^+-k_j}) \}
\cr &
-{1\over 4\beta }q
\sum_{j=1}^N\big({2(k_j^{(1)})^2\over k_i^+-k_j} -{(k_j^{(1)})^2\over
k_i^+ -k_j + \b m}- {(k_j^{(1)})^2\over k_i^+ -k_j - \b m}\big)\cr
&\quad 
&(4.30)
\cr}
$$
It remains to express all terms solely as a function of the quantum 
order parameters $a_i$. This is achieved by expanding $k_i ^+$ according to
(4.10) and recombining according to (4.21). The result may be simplified with
the help of the following two identities, valid up to ${\cal O}(q)$
$$
-\sum_{j\not=i}{1\over a_i-a_j}+
\2\sum_{j=1}^N\big[{1\over a_i-a_j+m}+{1\over a_i-a_j-m}\big]
=\beta {k_i^{(2)}\over (k_i^{(1)})^2}
\eqno(4.31)
$$
and
$$
\eqalignno{
{\p\over\p a_i}\sum_{j\not=i}(k_j^{(1)})^2
=&
{\p\over\p a_i}4 \beta ^2
\sum_{j\not=i}{\prod_{l=1}^N(a_j-a_{l}+m)(a_j-a_{l}-m)
\over \prod_{l\not=j}(a_j-a_{l})^2}\cr
=&
-\sum_{j\not=i}
({(k_j^{(1)})^2\over a_i-a_j+m}+{(k_j^{(1)})^2\over a_i-a_j-m}
-2{(k_j^{(1)})^2\over a_j-a_i})
&(4.32)
\cr}
$$
We can now combine the equations (4.30), (4.31), and (4.32), and obtain 
$$
\eqalignno{2 \pi i a_{Di}=
&a_i\L q +\2 m \L q\cr
&\quad- \sum_{j=1}^N\big[2(a_i-a_j)\L(a_i-a_j)
-(a_i-a_j+m)\L(a_i-a_j+m)\cr
&\quad\quad-(a_i-a_j-m)\L(a_i-a_j-m)\big]
+{1\over 4 \beta ^2}q{\p\over\p a_i}\sum_{j=1}^N(k_j^{(1)})^2
&(4.33)
\cr}
$$
We note that to order ${\cal O}(q)$, we may replace $k_i^+$
by $ \beta a_i$ in the expression (4.11) for $k_i^{(1)}$. Thus we arrive 
at the following final formula for $a_{Di}$ expressed entirely in terms of
the quantum order parameters $a_i$
$$
\eqalignno{2 \pi i a_{Di}=
&a_i\L q+\2 m\L q
-\sum_{j=1}^N\big[ 
2(a_i-a_j)\L(a_i-a_j) \cr
& \qquad -(a_i-a_j+m)\L(a_i-a_j+m)
-(a_i-a_j-m)\L(a_i-a_j-m)\big]\cr
&\qquad -q m^2 {\p\over\p a_i}\sum_{j=1}^N\prod_{l\not=j}
(1-{m^2\over (a_l-a_j)^2}).
&(4.34)}
$$
Notice that the dependence on the parameter $\beta$, which is related
to the $A$-period of $\Sigma$ and which can be chosen at will, has completely
disappeared from the above expression, as expected.

\bigskip

\noindent
{\bf  (d) The Prepotential}

\medskip

The expression for the prepotential is obtained by integration, and may be 
separated into a classical part $\F ^{{\rm class}}$, a perturbative part $\F
^{{\rm pert}}$, which arises only from one loop order, and an $n$-instanton
part $\F^{(n)}$.
$$
\F = \F ^{{\rm class}} + \F ^{{\rm pert}} + \sum _{n=1} ^\infty \F ^{(n)}
\eqno (4.35)
$$
We need at this point to make a distinction between
the prepotential $\F$, as determined by the original Seiberg-Witten
differential $d\l=kdz$, and its modification $\tilde \F$,
which is determined rather by the differential
$d\tilde\l=d\l+\2\b mdz$ with which we have worked
in Section IV.(c). We denote 
by $\tilde a_i$ and $\tilde a_{Di}$
the quantum
order parameters corresponding to $d\tilde\l$. 
From 
(4.34), the various contributions (4.35) to $\tilde\F$ are 
easily identified, up to
order ${\cal O}(q)$ included. We find
$$
\eqalignno{
\tilde\F ^{{\rm class}} =&  \2{\tau} \sum _{i=1} ^N \tilde a_i ^2
+\2 m \tau \sum _{i=1}^N\tilde a_i\cr
\tilde\F ^{{\rm pert}} = & -{1 \over 8 \pi i} \sum _{i,j =1} ^N \{
(\tilde a_i - \tilde a_j )^2 \log (\tilde
a_i - \tilde a_j )^2 -(\tilde a_i - \tilde a_j +m )^2 \log (\tilde
a_i - \tilde a_j+m )^2\}
\cr
\tilde\F ^{(1)} = & - {1 \over 2 \pi i} qm^2 \sum _{i=1} ^N \prod _{j\not=i} \big ( 1 -
{m^2 \over (\tilde a_i - \tilde a_j )^2} \big )
&(4.36) 
\cr}
$$
Now the relation between the order parameters $a_i$, $a_{Di}$,
and their modified version $\tilde a_i$, $\tilde a_{Di}$ is
$$
a_i=\tilde a_i-\2 m,\qquad a_{Di}=\tilde a_{Di}-\2m\t.\eqno(4.37)
$$
This implies the following relation between the prepotential
$\F$ and its modified version $\tilde\F$
$$
\F(a)=\tilde\F(a+\2 m)-\2 m\t\sum_{i=1}^Na_i.\eqno(4.38)
$$
(Although the term $\sum_{i=1}^Na_i$ is physically immaterial
(the Wilson effective action depends only on 
${\p^2{\cal F}\over\p a_i\p a_j}$),
we do not drop it at this point. This is in order to compare eventually
the present formulae with those obtained later from a renormalization
group equation. Also the
$SU(N)$ traceless constraint $\sum_{i=1}^Na_i=0$
is imposed only {\it after} differentiating $\F$ in $a_i$.)
Thus we obtain the following expression for $\F$
$$
\eqalignno{
\F ^{{\rm class}} =&  \2{\tau } \sum _{i=1} ^N a_i ^2+\2m\tau
\sum_{i=1}^Na_i\cr
\F ^{{\rm pert}} = & -{1 \over 8 \pi i} \sum _{i,j =1} ^N \{
(a_i - a_j )^2 \log (a_i - a_j )^2 -(a_i - a_j +m )^2 \log (a_i - a_j+m )^2\}
\cr
\F ^{(1)} = & - {1 \over 2 \pi i} qm^2 \sum _{i=1} ^N 
\prod _{j\not=i} \big ( 1 -
{m^2 \over (a_i - a_j )^2} \big )
&(4.39) 
\cr}
$$
where we have ignored an additional term
${3\over 8}m^2\tau N$, since it is $a_i$-independent.
Thus the prepotential $\F$ retains its form under a shift
of $d\l$ by a multiple of the form $dz$, which is
another reflection of the invariance of the
Lax equation $\dot L=[M,L]$ under a shift of $L$ by
a multiple of the identity. 
The contribution $\F ^{{\rm class}}$ agrees with the well-known 
form at tree
level. The contribution $\F ^{{\rm pert}}$ is precisely the one expected 
for a system consisting of gauge multiplet states with masses $|a_i-a_j|$
and of hypermultiplet states with masses $|a_i-a_j+m|$. Notice that both
contributions enter with opposite signs as expected. Finally, the
contribution $\F^{(1)}$ has not as yet been evaluated starting from
conventional field theory methods. It would be interesting to compare our
answer with a direct instanton calculation. In Section VI, we present
various decoupling limits of the theory and check in particular that 
$\N=2$ supersymmetric pure $SU(N)$ gauge theory is recovered upon
decoupling the hypermultiplet by letting $m \to \infty$, and suitably
adjusting the coupling $\tau$.

\vfill\break

\centerline{\bf V. THE RENORMALIZATION GROUP EQUATION}

\bigskip

In this section, we shall derive a renormalization group equation
for the $\N=2$ supersymmetric $SU(N)$ gauge theory with matter in the adjoint
representation. Although the techniques of [11] and [15-18] (see also [19])
were successful
in the case of asymptotically free theories
with matter in the fundamental representation, they
yield in the present case only the homogeneity relation
$$
\sum_{i=1}^N a_i{\p{\cal F}\over\p a_i}+m{\p{\cal F}\over\p m}
-2{\cal F}=0.
\eqno(5.1)
$$
A much more powerful equation may be produced by considering a
renormalization group equation for the variation of $\F$ with respect to
$\tau$.

\bigskip

\noindent
{\bf (a) Derivation of the Renormalization Group Equation} 

\medskip

The key starting ingredient for our derivation is the following
variational formula for the $B_i$-periods $a_{Di}$ of the Seiberg-Witten
differential $d\lambda = \tilde k dz$, as defined in (3.14).
$$
\eqalign{
\delta a_{Di}=&
{1\over 2\pi i}\int\int_{\G}\mu  \tilde k \,d\bar z\wedge d\omega_i 
+\sum_{\a=1}^N M_\a V(P_{\a})\phi_i(P_{\a}) \cr
 M_1 = & -(N-1) m \cr
 M_\a =& \ m, \ \ \ \a = 2, \cdots , N 
\cr}
\eqno (5.2)
$$
where 
$\mu=\mu_{\bar z}^{\ z}d\bar z\otimes{d\over dz}$ is the 
Beltrami differential on $\G$ induced
by a moduli deformation $\t\to \t+\delta \t$ of the base
torus $\S$. Here we have represented $\mu$ as $\mu=\p_{\bar z}V(z)$,
with $V$ a vector field with jump discontinuties
across the cycles $A_i$. We have also introduced the basis
$d\omega_i$, $i=1,\cdots,N$, of
holomorphic Abelian differentials for the
surface $\Gamma$ which is dual to the cycles $A_i$, and 
the associated functions $\phi_i(z)$ by
$d\omega_i=\phi_i dz$.

\medskip

To establish (5.2), we note that the
Seiberg-Witten differential
in the case of matter in the adjoint representation satisfies
the equation
$$
\p_{\bar z}d\l=-\pi\sum_{i=1}^N M_\a \delta (z,P_{\a}).
\eqno(5.3)
$$
To vary this equation with respect to $\tau$, we recall that
the $\p _{\bar z}$ operator on scalars and on one-forms
varies respectively by $-\mu\p_z$ and $-\p_z\mu$,
under
a deformation of the complex structure by a Beltrami
differential $\mu$ [20,21]. This implies that the
local coordinate $z$ is deformed to $z+V(z)$. Thus
the variational equation derived from (5.3) is
$$
\p_{\bar z}\delta(d\l)
=-\p_z(\mu d\lambda)
+\pi\sum_{\a=1}^N M _\a \{ 
  V(P_{\a})\p_{P_{\a}}\delta(z,P_{\a})
+ \bar V(P_{\a})\p_{\bar P_{\a}}\delta(z,P_{\a}) \}.
\eqno (5.4)
$$
Let $E(z,w)$ be the prime form. Since the Green's function for 
the operator $\p_{\bar z}$ on 
(1,0) forms is the Szeg\"o kernel ${1\over\pi}\p_z\L E(z,w)$, 
we obtain
$$
\eqalignno{ 
\delta(d\lambda)(z)
=&{1\over 2\pi i}\int\int_{\G}
\p_w\p_z\L E(z,w)(\mu  \tilde k(w))d\bar w\wedge d w\cr
&
+\sum_{\a=1}^N M_\a V(P_{\a})\p_{P_{\a}}\p_z\L E(z,P_{\a})
+\pi\sum_{\a=1}^N M_\a \bar V (P_\a) \delta(z,P_{\a}).
&(5.5)\cr}
$$
Now the prime form satisfies the following
identities [22]
$$
\eqalignno{
\oint_{A_i}dz\,\p_z\p_w\L E(z,w)=&0\cr
\oint_{B_i}dz\,\p_z\p_w\L E(z,w)=&2\pi i\phi_i(w).
&(5.6)\cr}
$$
Thus integrating the right hand side of (5.5)
over the $A_i$ cycles shows that it
has zero $A$-periods (as it should have), while
integrating over the $B_i$ cycles gives the desired equation (5.2),
in view of the fact that the punctures $P_{\a}$ do not
lie on the $B_i$ cycles. 

So far, our considerations
have been quite general, and would apply with little change
to any deformation of the moduli of $\G$. In the
present situation, the deformation is of the base
torus, and thus the vector field $V(z)$
has the same constant discontinuity across all the
$A_i$ cycles. We may rewrite the surface integral
in the right hand side of (5.2) as
$$
{1\over 2\pi i}
\int\int_{\G}\mu  \tilde k d\bar w\wedge d\omega_i
={1\over 2\pi i}\lim_{\epsilon\to 0}\int\int_{\G\setminus
\cup_{\a=1}^N\{|z-P_{\a}|<\epsilon\}}d(V \tilde k\,d\omega_i).
$$
In view of Stokes' theorem, this can be
re-expressed as a the difference of a line
integral around the boundary $\prod_{i=1}^NA_iB_iA_i^{-1}B_i^{-1}$
of the cut surface, and simple contour integrals
around the poles $P_{\a}$ of $\tilde k$. Only the poles of $\tilde k$
and the discontinuities $V^+ -V^-$
of $V$ across the $A_i$ cycles contribute, since $V$ is continuous across the
$B_i$ cycles and the contributions
of $B_i$ and $B_i^{-1}$ cancel. Thus
$$
{1\over 2\pi i}
\int\int_{\G}\mu  \tilde k d\bar w \wedge d\omega_i
= {1\over 2\pi i}(V^+-V^-) 
\sum _ {j=1}^N\oint_{A_j} \tilde k \, d\omega_i
-
\sum_{\a=1}^N M_\a V(P_{\a}) \phi _i (P _\alpha)
$$
and (5.2) reduces to
$$
\delta({1\over 2\pi i}\oint_{B_i}d\l)
={1\over 2\pi i}(V^+-V^-)\sum_{j=1}^N\oint_{A_j}
\tilde k \,d\omega_i.
\eqno(5.7)
$$
We can determine the correct value of the jump
$V^+-V^-$ by comparing the preceding formula
with the case of moduli
deformations of the base torus $\S$. In fact,
the above argument applies equally well to
the variations of the differential $dz$
on the base $\S$ in place of the variations
of $d\l$ on the surface $\G$.
We obtain in this case
$$
\delta(\oint_Bdz)=(V^+-V^-)\oint_Ad\omega 
\eqno(5.8)
$$
where $A$ and $B$ are now cycles on $\S$, and $d\omega$
is the Abelian differential dual to the $A$ cycle.
Varying $\tau$ with, say, the length $2\omega_1$
of the $A$ cycle fixed, we find
$$
V^+-V^-=-2\omega_1 \delta \tau
\eqno(5.9)
$$
where the - sign is due to the fact that,
with the orientation of the $A_i$ and $B_i$ cycles
described at the beginning of Section IV,
the $B$ cycle actually goes from $\omega_2$ to
$-\omega_2$ in the fundamental domain for
$\S={\bf C}/2\omega_1{\bf Z}+2\omega_2{\bf Z}$.
The equation (5.7) becomes
$$
{\partial a_{Di} \over \partial \tau}
=-{2\omega_1\over 2\pi i}
\sum_{j=1}^N\oint_{A_j} \tilde k d\omega_i
\eqno(5.10)
$$

\medskip

We can reexpress the equation
(5.10) in terms of the prepotential
${\cal F}$. Since the poles of $d\l$ are independent
of the order parameters $a_i$, the derivatives
of $d\l$ with respect to $a_i$ must be holomorphic Abelian
differentials. Since the $a_i$ are 
the $A_i$ periods of $d\l$, we have
$$
d\omega_i={1\over 2\pi i}{\p\over\p a_i}d\l
={1\over 2\pi i}{\p  \tilde k \over\p a_i}dz.
\eqno(5.11)
$$
In terms of ${\cal F}$, the equation (5.10) is then
$$
{\p\over\p\t}({\p{\cal F}\over\p a_i})={2\omega_1\over 8\pi^2}
{\p\over\p a_i}\big(\sum_{j=1}^N
\oint_{A_j} \tilde k^2dz\big).
\eqno(5.12)
$$
This identifies the renormalization group equation
for the $\N=2$ supersymmetric gauge theory
with matter in the adjoint representation to be
$$
{\p{\cal F}\over\p\t}
={2\omega_1\over 8\pi^2}\sum_{j=1}^N
\oint_{A_j}\tilde k^2dz.
\eqno(5.13)
$$
We can view the right hand side of (5.13) as an {\rm exact}
formula for the beta function of the theory. It is
essentially given by the Hamiltonian
of the associated integrable system. Indeed, the summation over all
cycles $A_j$ of (5.13) can be replaced by the integral over
the single cycle $A$ downstairs of $\tilde k_1^2+\cdots+\tilde k_N^2=Tr L^2$,
where the $\tilde k_i$, $i=1,\cdots,N$, denote
the $N$ roots of the defining equation $\det(\tilde kI-L(z))=0$
of the spectral curve. Explicitly, we have
$$
{\p{\cal F}\over\p\t}
={2\omega_1\over 8\pi^2}\oint_A Tr L^2\,dz
=-{\omega_1^2\over 2\pi^2}\left(\sum_{i=1}^Np_i^2
-m^2\sum_{i\not=j}\wp(x_i-x_j)\right).
\eqno(5.14)
$$
where the $a_i$-independent term
${\omega_1\eta_1\over2\pi^2}N(N-1)$ has been dropped.
Finally, for computational purposes,
it is useful to recast the RG equation (5.13) in terms
of the variable  $k$ in $f(k - \2 \beta m,z)=0$ rather than the original
variable $\tilde k$ in $R( \tilde k,z)=0$. Recall that
$k =  \tilde k + m h_1(z)  + \2 \beta m $,
and thus
$$
k^2=\tilde k^2+2\tilde k(mh_1(z)+\2\b m),
$$
where we have ignored the term $(mh_1+\2\b m)^2$, since it contributes
only an $a_i$-independent term to the prepotential.
The integral over the sum of all $A_j$ cycles of the second
term on the right hand side of the above equation can be again replaced
by an integral over the cycle $A$ downstairs,
with $\tilde k$ replaced by the trace $\sum_{i=1}^Np_i$
of the matrix $L(z)$ (c.f. (2.11)). We obtain
$$
\sum_{j=1}^N\oint_{A_j}
2\tilde k(mh_1(z)+\2\b m)dz=
2(\sum_{j=1}^Np_j)\oint_A
(mh_1(z)+\2\b m)dz=0,
$$
in view of the transformation law (3.3) for the $\vartheta_1$-function,
and the fact that $h_1(z)=\p_z\L\vartheta_1({z\over 2\omega_1}|\t)$.
Thus we may use either $\tilde k^2$ or $k^2$ in the expression
(5.13) for the beta function. 

\medskip

As an immediate check, we determine
the classical part $\F^{\rm class}$ of the
prepotential (4.42) from the renormalization
group equation, leaving the more interesting derivation
of the one-instanton and two-instanton contributions to
the next section. Ignoring then ${\cal O}(q)$ terms,
we find the following value for the beta function,
in view of the expansion (4.13) and (4.19) for $dz$
$$
{\p\F\over\p\t}={1\over 2\beta^2}\sum_{i=1}^Nk_i^2 +{\cal O}(q).
$$
Since (4.21) implies that $\beta^{-1}k_i=a_i+\2 m+ {\cal O}(q)$
(recall that the left hand side of (4.21) is actually
$\tilde a_i$, and
that $a_i$ and $\tilde a_i$ are related by (4.37)),
we obtain
$$
{\p\F\over\p\t}
=\2\sum_{i=1}^Na_i^2+\2 m\sum_{i=1}^Na_i+{\cal O}(q),
$$
in agreement with (4.39).

\medskip

Henceforth, it is convenient to set $\beta =1$, i.e. $ \omega _1 = - i
\pi$, since the prepotential is independent of $\beta$ anyway. Our
final renormalization group equation takes the form
$$
{\p{\cal F}\over\p\t}
={1\over 4\pi i}\sum_{j=1}^N
\oint_{A_j} \tilde k^2dz={1\over 4\pi i}\sum_{j=1}^N
\oint_{A_j} k^2dz.
\eqno(5.15)
$$
The power of this formula lies in the fact that all terms in an expansion in
powers of $q$ may be evaluated using residue methods only, just as was the 
case for the calculation of the $A_j$ periods. To show how this works, we
produce now a calculation of the 1- and 2-instanton contributions.

\bigskip

\noindent
{\bf (b) 1- and 2- Instanton Results from the Renormalization Group Equation}

\medskip

In this section, we evaluate the 1- and 2-instanton contributions to the
prepotential, using the renormalization group equation of (5.15). As stated
there, we
set $\beta =1$. The only objects we need to calculate are the $a_i$ quantum
order parameters, as integrals over $A_i$ cycles of the differential $kdz$, 
and the RG beta function in (5.15) as integrals over $A_i$ cycles of the
differential $k^2 dz$. Both calculations are carried out by residue methods
only ! As in Section IV, we choose $k$ as independent variable, and need to
express $dz = d \log w$ as a function of $k$. This is done with the help of
(4.13), and the expansion of $\log y$ is obtained from (4.14) and (4.15). 
Since the $A_i$ cycles may be chosen away from the poles $k_i$ by a 
distance of order ${\cal O}(q^0)$, it suffices to use an expansion for $\log
w$ or $\log y$ in powers of $q$, and for $k-k_i$ or order ${\cal O}(q^0)$.
In particular, we do not need to worry about $k$ coming close to $k_i$ by a
distance of order ${\cal O}(q^\2)$, as we had to in equations (4.17) to
(4.19). Taking these considerations into account, we find
$$
\log y = q(-\eta _1  + \bar \eta _1) + \2 q^2 (-3 \eta _1 ^2  + 3 \bar \eta
_1 ^2) + {\cal O}(q^3)
\eqno (5.16)
$$
where 
$$
\eta _1 = {H(k+m) H(k-m) \over H(k)^2}
\qquad
\bar \eta _1 = {H(k) H(k-2m) \over H(k-m)^2}
\eqno (5.17)
$$
The differential $dz$ thus becomes
$$
dz = d \log H(k) - d \log H(k-m) + q(- \eta _1 ' + \bar \eta _1 ') dk
+q^2 (-3 \eta _1 \eta _1 ' + 3 \bar \eta _1 \bar \eta _1 ') dk.
\eqno (5.18)
$$
where the prime stands for derivation with respect to $k$.
Now, in evaluating the A-periods using residue methods at the zeros of 
$H(k)$, the contribution $d \log H(k-m)$ and terms involving $\bar \eta _1$
never enter since they do not exhibit poles at $k=k_i$. Thus, we are left
with dependence only on $H(k)$ and on the function $\eta _1$. Residue
calculations at the poles
$k_i$ will in fact only involve the functions
$$
\bar S_i (k) \equiv {H(k+m) H(k-m) \over H_i (k)^2}.
\eqno (5.19)
$$ 
(Notice that with this notation, we have $(k_i ^{(1)}) ^2 = 2\bar S_i (k_i)$ 
and $k_i ^{(2)} = 2 \bar S_i '(k_i)$, according to (4.11).) We find, again to
order ${\cal O}(q^2)$ : 
$$
\eqalign{
a_i = & k_i + q \bar S_i '(k_i) + {1 \over 4} q^2 \{ \bar S_i ^2\} ''' (k_i)
 \cr
{\p \F \over \p \tau} = & 
 \sum _{i=1} ^N \bigg \{ \2 k_i ^2 +  q k_i \bar S_i ' (k_i) 
+ q \bar S_i (k_i)
+{1 \over 4} q^2 k_i \{ \bar S_i ^2\} ''' (k_i)  
+{ 3\over 4}  q^2 \{ \bar S _i
^2\} '' (k_i) \bigg \}
\cr}
\eqno (5.20)
$$
where the prime denotes taking a derivative with respect to $k$ and setting
$k=k_i$ afterwards. It now simply remains to recast the second equation in
(5.20) in terms of variable $a_i$ only, and this is achieved using the first
equation in (5.20). It is convenient to define the function $S_i(a)$, 
which is the analogue of $\bar S_i (k)$, but with all variables $k_i$
replaced by $a_i$; explicitly, we have
$$
S_i(a) =  {\prod _{j=1} ^N (a-a_j+m) (a-a_j -m) \over \prod _{j\not=i} ^N
(a-a_j)^2}
\eqno (5.21)
$$
In terms of this function, we find
$$
\eqalign{
{\p \F \over \p \tau} = & 
 \sum _{i=1} ^N \bigg \{ 
\2 a_i ^2 +  q  S_i (a_i) + q^2  S _i ' (a_i) ^2 
+{3 \over 2}  q^2 S_i(a_i) S_i
'' (a_i)
 \bigg \}
\cr
&\ - q^2 \sum _{i,j=1} ^N S_j ' (a_j) {\p \over \p a_j} S_i (a_i)
\cr}
\eqno (5.22)
$$
which may be integrated with respect to $\tau$ (keeping $a_i$ fixed) in a
straigthforward way. We find
$$
\eqalign{
 \F  =  \F ^{{\rm class}} + \F^{{\rm pert}}  & +
{1 \over 2 \pi i} \sum _{i=1} ^N \bigg \{ 
 q  S_i (a_i) + \2 q^2  S _i ' (a_i) ^2 + {3\over 4} q^2 S_i(a_i) S_i ''
(a_i)
 \bigg \}
\cr
& -{1 \over 4 \pi i} q^2 \sum _{i,j=1} ^N S_j ' (a_j) {\p \over \p a_j} S_i
(a_i)
\cr}
\eqno (5.22)
$$
It will be convenient for our study of various decouplings in Section VI, to
recast (5.22)  by rearranging the 2-instanton contributions in the following
form.
$$
\eqalign{
 \F  =  \F ^{{\rm class}}  & + \F^{{\rm pert}}   +
{1 \over 2 \pi i} \sum _{i=1} ^N \bigg \{ 
 q  S_i (a_i)  + {1\over 4} q^2 S_i(a_i){\p ^2  S_i  (a_i) \over \p a_i ^2}
 \bigg \}
\cr
& +{1 \over 2 \pi i} q^2 \sum _{i\not= j=1} ^N S_i (a_i) S_j  (a_j) 
\bigg [ { 1 \over (a_i -a_j)^2} 
       - \2 {1 \over a_i-a_j +m}  - \2 {1 \over a_i-a_j -m} \bigg ].
\cr}
\eqno (5.23)
$$
 It is easy to see
that the classical part is that of (4.39), and that the 1-instanton
contribution (linear in $q$) also agrees with (4.39). The perturbative part
$\F ^{{\rm pert}}$ is independent of
$\tau$ and cannot be obtained from the renormalization group equation; it was
taken over from (4.39). The 2-instanton contribution (quadratic in $q$) is
new.   In Section VI, we shall show that upon decoupling the hypermultiplet
by sending $m \to \infty$, (5.23) converges to the prepotential for the pure
$\N=2$  theory  with gauge group $SU(N)$, as obtained in [7,11].

\bigskip
\bigskip 

\centerline{{\bf VI. DECOUPLING LIMITS AND PRODUCT GAUGE GROUPS}}

\bigskip

The independent parameters in the $\N=2$ supersymmetric $SU(N)$ gauge theory
are the complex gauge coupling $\tau$, the hypermultiplet mass parameter $m$,
and the quantum order parameters $a_i$, $i=1, \cdots, N$ (or equivalently the
classical order parameters $k_i$). 
The masses of the gauge multiplet and hypermultiplet states are respectively
given by $|a_i - a_j| $ and $|a _i - a_j + m|$. By taking various singular
limits of the parameters, subsets of the states of the theory may be given
infinitely large mass and made to decouple. What remains after decoupling 
is a different $\N=2$ supersymmetric gauge theory. To be more precise,
singular limits of the parameters may induce the following two effects.

\medskip

\item{(1)} States whose mass tends to $\infty$ disappear from the spectrum 
and decouple from the dynamics of the theory. The gauge group of the 
remaining theory is in general a subgroup (which need not be a simple group)
of the original gauge group $SU(N)$.
\medskip
\item{(2)} The dynamics associated to one or several of the gauge subgroups 
may freeze out when the effective coupling of this gauge subgroup tends to
zero. The gauge states of the corresponding gauge subgroup become
non-interacting, and the adjoint scalar fields that belong to the $\N=2$
multiplet of the gauge subgroup are frozen to their vacuum expectation
values.

\medskip

In this section, we shall describe various decoupling limits of the $\N=2$
theory with a massive adjoint hypermultiplet. 
We begin by disposing of some cases which are uninteresting. First, keeping
$m$ and $a_i$ finite and letting $\tau \to \infty$ produces a free theory.
Second, keeping $\tau $ and $m$ finite, and letting some $a_i \to \infty$,
produces an $\N=2$ gauge theory with an adjoint hypermultiplet and a gauge
group which is a product $SU(N_1) \times \cdots \times SU(N_p)$ subgroup 
(and possibly U(1) factors) of $SU(N)$, with decoupled dynamics between
different factors. Third, keeping $m$  finite, and letting $\tau \to \infty$
and some of the $a_i \to \infty$, we recover again a free theory. 

\medskip

Thus, to obtain interesting decoupling limits, we must let $\tau \to \infty
$ and at the same time $m \to \infty $ in a related way. Notice that in
this case either the gauge mass $|a_i - a_j|$ or the hypermultiplet mass
$|a_i - a_j +m|$ must tend to $\infty$, since both could not be kept finite
at the same time. Thus for given $ij$, either the gauge or the
hypermultiplet state must decouple.  We distinguish two cases :
\medskip 
\item{(a)} All $a_i$ remain finite, in which case the gauge group remains
$SU(N)$, but the full hypermultiplet decouples. We show in subsection (a)
below that the pure $SU(N)$ theory is indeed recovered. This decoupling limit
was also exhibited in [2] using the curve obtained through the Hitchin
system.

\medskip

\item{(b)} Some $a_i$ are also sent to $\infty$ in such a way that certain
hypermultiplet masses remain finite. In this limit, the gauge group
$SU(N)$ is broken to a subgroup of the type $SU(N_1) \times \cdots \times
SU(N_p)$. (We shall see that $U(1)$ factors, which in principle could appear,
actually always decouple.) The remaining hypermultiplets may be in 
fundamental representations of one of the gauge group factors or in
bi-fundamental representations of two of the gauge factors (not all such
bi-fundamentals are allowed though). In subsection (b) below, we analyze
this case when two factors arise (this includes the case where the dynamics
of one of the factors freezes out as explained in (2) above.) In subsection
(c) we analyze the general case, and discuss as an application the case
of 3 factors in (d).  
 
\medskip

\noindent
We shall now examine each of these cases in turn.

\bigskip

\noindent
{\bf (a) Decoupling the full hypermultiplet}

\medskip

The pure $\N=2$ supersymmetric $SU(N)$ gauge theory is recovered by 
decoupling the full hypermultiplet in the limit where $\tau \to \infty$, $m
\to \infty$ while keeping constant the parameters $a_i$ and $\Lambda$ :
$$
\Lambda ^{2N} = (-)^N m^{2N} q
\qquad \qquad 
q= e^{2 \pi i \tau}.
\eqno (6.1)
$$
Notice that since $q \to 0$ in this limit, it is equivalent to keep the
classical order parameters $k_i$ fixed. The curve of (4.5) then converges to the
curve of the pure theory upon scaling the variable $w$ in such a way that 
$t$, as defined below, is kept fixed
$$
H(k) - t - { 1 \over t} \Lambda ^{2N}=0
\qquad \qquad
w= t (-m)^{-N} .
\eqno (6.2)
$$
The Seiberg-Witten differential directly follows from the same change of 
variable $z =\log w$ to $t$ in (6.2) : $d\lambda = k d\log t$. 

The limit of the effective prepotential in (5.23) is readily obtained by
first establishing that 
$$
qS_i (a_i) \to \Lambda ^{2N} \prod _{i \not= j} ^N {1 \over (a_i - a_j)^2}.
\eqno (6.3)
$$ 
and is found to agree (up to an irrelevant additive constant) with 
expression (4.34a,b,c) of [7] for the pure case $N_f=0$.

\medskip

Next, we show that the renormalization group equation of (5.15) reduces to 
the renormalization group equation for the case of the pure theory. As the
curve for the adjoint case converges to that of the pure case (6.2), the
curve becomes hyperelliptic. The sum over the $A_i$-cycles in (5.15) may be
deformed into a single contour encircling all $A_i$ branch cuts, which in
turn may be deformed into a contour around $k=\infty$. The sum over the
corresponding contour integrals may now be evaluated by residue methods
around $k=\infty$. Upon defining $s_2$ by $H(k) = k^N + s_2 k^{N-2} + {\cal
O}(k^{N-3})$, and using the relation (6.1) to convert the variation with
respect to $\tau$ into a variation with respect to $\Lambda$ while keeping
$m$ fixed, we find
$$
{\p \F \over \p \log \Lambda} = -{2 N \over 2 \pi i} s_2
\eqno (6.4)
$$
in agreement with equation (3.20) of [11].

Finally, we note that the decoupling of the full hypermultiplet may also be
carried out directly on the Lax operator $L(z)$ in (2.11), by taking the
well-known singular limit of the Calogero-Moser system to the affine Toda
system [23] for group $SU(N)$. In this way, we recover the curve for the
pure
$SU(N)$ theory, obtained as the spectral curve of the Lax operator for affine
Toda for $SU(N)$ in [24].

\bigskip

\noindent
{\bf (b) Decoupling $ SU(N) \to SU(N_1) \times SU(N_2)$}

\medskip

Next, we consider a singular limit, in which the quantum order parameters 
$a_i$ (or equivalently the classical order parameters $k_i$)
fall into two groups with $i=1, \cdots , N_1$ and $j= N_1 +1 ,\cdots , N=N_1
+N_2$, (with $N_1, N_2 \geq 1$). The groups are such that as $m \to\infty$, 
we have $(a_i -a_j)/m\to 0$ when $i,j$ belong to the same group and $(a_i -
a_j)/m\to \pm 1$ when $i,j$ belong to different groups. It is clear that the
light gauge states fill out a multiplet of the gauge group $SU(N_1) \times
SU(N_2) \times U(1)$, and that the light hypermultiplet states transform 
under the bi-fundamental representation $({{\bf N}_1}, \overline {{\bf
N}}_2)$ $
\oplus$ $(\overline {{\bf N}} _1,{{\bf N}}_2) $ of the semi-simple part of 
the gauge group. 

\medskip

The curve and effective prepotential may be obtained as a limit of the case 
with adjoint hypermultiplet. However, special care is required in the
analysis of the three gauge couplings of the gauge group $SU(N_1) \times
SU(N_2) \times U(1)$. Indeed, one or several of the couplings may flow to
zero in the limit and the dynamics of the corresponding gauge group may
freeze out, as discussed in (2) above. In fact, this will always be the case
for the $U(1)$ part of the gauge group, but may also occur for one of the
simple factors.

\medskip

To analyze the singular limits quantitatively, we decompose the classical
(equivalently quantum) order parameters in terms of order parameters $x_i$
and $y_j$ of the gauge groups $SU(N_1)$ and $SU(N_2)$ respectively 
$$
\eqalign{
k_i = & v _1 + x_i 
\qquad i = 1, \cdots , N_1 \cr
k_{N_1 +j} = & v_2 +  y_j 
\qquad j = 1, \cdots , N_2. \cr}
\eqno (6.5)
$$
Uniqueness of the decomposition is achieved by imposing the conditions $\sum
_i x_i = \sum _j y_j=0$, so that $N_1 v_1 + N_2 v_2 =0$. The order parameter
$v\equiv v_1 - v_2$ is associated with the $U(1)$ factor of the gauge group. 

\medskip

\noindent
{\it The Limiting Curve}

We now assume that $x_i$, $y_j$ and $\mu = v-m$ are kept fixed in the limit, 
and determine the behavior of $\tau$ and $m$ (which tend to $\infty$) that
can yield interesting limits. First, we determine the limiting curves and
then analyze which gauge couplings tend to zero. The limit of the curve
(4.5) for adjoint hypermultiplet is obtained by deriving the large $m$
behavior of the polynomial $H(k)$. It is convenient to work with the
variable $x=k-v_1$, since it has a finite limit as $m \to \infty$. 
$$
H(k) = \prod _{i=1} ^{N_1} (x - x_i ) 
       \prod _{j=1} ^{N_2} (x +m + \mu -y_j),
\qquad k=x + v_1.
\eqno (6.6)
$$
We define the polynomials 
$$
\eqalign{
A(x) \equiv &  \prod _{i=1} ^{N_1} (x-x_i) \cr
B(x) \equiv & \prod _{j=1} ^{N_2} (x+ \mu - y_j), \cr }
\eqno (6.7)
$$
and obtain the leading large $m$ behavior of $H(k)$ as a function of $A(x)$ 
and $B(x)$ :
$$
\eqalign{
H(k)      \sim & \ m^{N_2} A(x) \cr
H(k-m)    \sim & \ (-m)^{N_1} B(x) \cr
H(k - nm) \sim & \ c_n m^{N_1 + N_2}, 
\qquad c_n = (-n)^{N_1} (-n+1)^{N_2} \quad  \ n\not=0,1. \cr}
\eqno (6.8)
$$
It is convenient to redefine $w$ in terms of a parameter 
$t$ that has a finite
limit as $m\to \infty$ :
$$
w\equiv t  m^{N_2 - N_1}.
\eqno (6.9)
$$
The leading large $m$ behavior of the curve is now given by
$$
 A(x) - t (-) ^{N_1} B(x) 
+ \sum _{n\in {\bf Z},n\not=0,1}  (-)^n q ^{ \2 n (n-1)}
 t^{n} c_{n}  m^{\nu _{n}} =0
\eqno (6.10)
$$
where $\nu _n = N_1 + n(N_2-N_1)$. 

\medskip

It remains to derive the behavior of $q$ as
a function of $m$ in the limit. To do so, we assume, without loss of
generality, that $N_2 \leq N_1$. In (6.10), the orders $n=-l$ and $n=l+1$
with $l
\geq 0$ enter with the same $q$-dependence, but, we have $\nu _{-l} - \nu
_{l+1}= -(2l+1) (N_2-N_1) \geq 0$. Thus, to order $q$, the term with $n=-1$
dominates in (6.10) over the term with $n=2$ if $N_2 < N_1$, while both 
terms are on the same order for $N_2 = N_1$. In either case, an interesting
limit is obtained by holding $\Lambda $ fixed in the following $q$
dependence 
$$
q  m ^{2N_1 - N_2} \equiv \Lambda ^{2N_1 - N_2}. 
\eqno (6.11)
$$
It is easy to see that with this dependence, all terms in (6.10) with  $n
\not= -1,0,1,2$ tend to 0.
The limiting curve is
$$
A(x) - t (-)^{N_1} B(x) 
 -  2^{N_2} \Lambda ^{2N_1 - N_2} \big [ {1 \over t}
- t^2 (-)^{N}
\bigl ( {m^3 \over 2} \bigr ) ^{N_2-N_1} \big ] =0
\eqno (6.12)
$$
Of course, when $N_2 < N_1$, the last term in (6.12) may be dropped.

\medskip

\noindent
{\it Flow of Couplings, and Freezing out of the $U(1)$}

To analyze the physical system governed by curve (6.12), we study the 
behavior of the three couplings, associated with the gauge groups $SU(N_1)$,
$SU(N_2)$ and $U(1)$ respectively. As the coupling $q$ tends to zero in the
limit of $m
\to \infty$, as given by (6.11), it suffices to use perturbation theory
to do so, retaining only $\F ^{{\rm class}} + \F ^{{\rm pert}} \equiv i \Phi
/8\pi $ in the expansion of the prepotential (4.36). To this order, the 
quantum order parameters $a_i$ in (4.36) may be replaced by the classical
order parameters $k_i$. In order to keep track of dimensionful parameters
in  a consistent way, it is convenient to introduce an {\it arbitrary} scale
parameter inside the logarithmic terms, which we shall choose to be $\Lambda
>0$, introduced in (6.11).
The function $\Phi$ splits up into a part with a manifestly finite limit 
$$
\eqalign{
\Phi ^{{\rm lim}} 
= &   \sum_{i,j=1}^{N_1}(x_i - x_j)^2 \log (x_i-x_j)^2/ \Lambda^2 
  + \sum_{i,j=1}^{N_2}(y_i - y_j)^2 \log (y_i-y_j)^2/ \Lambda^2 
\cr
  & \qquad  - \sum_{i=1}^{N_1} \sum _{j=1} ^{N_2} (x_i - y_j + \mu )^2 \log
   (x_i-y_j + \mu )^2/ \Lambda^2, 
\cr}
\eqno (6.13)
$$
a part which involves only quadratic dependence on $x_i$, $y_j$ and $\mu$
$$
\Phi ^{{\rm quad}} = C_x \sum _{i=1 } ^{N_1} x_i ^2 + C_y \sum _{j=1} ^{N_2}
y_j ^2 + C_\mu \mu ^2,
\eqno (6.14)
$$
and a part which is independent of $x_i$, $y_j$ and $\mu$ and which may thus
be dropped. The coefficients $C$ are easily computed and we find
$$
\eqalign{
C_x = & -4 \pi i \tau + (N_2-2N_1)(\log m^2/ \Lambda ^2 +3) - N_2 \log 4 
\cr
C_y = & -4 \pi i \tau + (N_1-2N_2)(\log m^2/ \Lambda ^2 +3) - N_1 \log 4 
\cr
C_\mu = & \{ - 4 \pi i \tau /N + \log m^2/4 \Lambda ^2 +3 \} N_1 N_2.
\cr }
\eqno (6.15)
$$
It is clear from (6.15) that as $m\to \infty$ and $\Im \tau \to  \infty$, we
will always have $C_\mu \to \infty$. This means that the $U(1)$ gauge
coupling tends to zero so that the dynamics of this part of the gauge group
freezes out. The field $\mu$ (which was the scalar component of the $\N=2$
vector multiplet associated with the $U(1)$ factor) freezes to a constant. 
Thus, the gauge group of the limiting theory is only $SU(N_1) \times SU(N_2)
$.

\medskip

\noindent
{\it The Case $N_1 = N_2$}

Next, we investigate under which conditions the effective couplings for both
$SU(N_1)$ and $SU(N_2)$ remain finite as $m,\ Im \tau \to \infty$. This will
require that both $C_x$ and $C_y$ converge to a finite limit. It is clear 
from (6.15) that {\it this can happen if and only if $N_1=N_2$}. In that
case, both couplings are the same, and the theory has $SU(N_1) \times
SU(N_1) $ gauge group with a hypermultiplet in the bi-fundamental
representation  $({{\bf N}_1}, \overline {{\bf N}}_1)$ $ \oplus$ $(\overline
{{\bf N}} _1,{{\bf N}}_1)
$, with mass $\mu$.

\medskip

The curve in this case is given by (6.12), and simplifies to
$$
A(x) - t (-)^{N_1} B(x)- 2^{N_1} \Lambda ^{N_1 } \big [ {1 \over t} -
t^2 \big] =0
\eqno (6.16)
$$
Its form is easily seen to coincide with the model for the product group
$SU(N_1) \times SU(N_1) $ solved by Witten using M-theory and D-brane
technology [12]. The same form was also derived by Katz, Mayr, and Vafa [25]
using compactifications of Type II strings 
on Calabi-Yau manifolds and the mirror symmetry of K3.

Although the curve (6.16) is relatively complicated (say, compared to the
curve (3.1) of the $SU(N)$ theory with matter in the fundamental
representation), its effective prepotential to any order of instanton
corrections can be easily read off from the corresponding
prepotential for the $SU(N)$ theory with matter in the adjoint
representation by taking the limit $m\to\infty$, $q\to 0$, 
$qm^{N_1}=\Lambda^{N_1}$ fixed. It is convenient to introduce the notation
$x_i=x_i^1$, $y_i=x_i^2$. For each $I$, the index $i$ in $x_i^I$
should be viewed as running over a range $1\leq i\leq N_I$, which we
denote more informally by ``$i\in I$". We set
$$
\eqalignno
{&A_i^I (x) = \prod _{j\not=i\atop j\in I} (x-x_j^I)
\qquad \qquad
B^I(x) = \prod _{j\in J\atop |I-J|=1} \big(\mu\pm (x - x_j^J)\big)\cr
&S_i^I(x)={B^I(x)\over A_i^I(x)^2},
&(6.17)\cr}
$$
where the sign $\pm$ in the expression
$B^I(x)$ is the same as the sign of
$J-I$. Then the limit of the function $S_i(a)$ in (5.21)
is given by
$$
q S_i(a_i) \to (-2\Lambda) ^{N_1} S_i^I(x_i^I).
$$
We may now substitute in the expansion (5.23) and obtain this way
the prepotential to two-instanton order for the
$SU(N)\times SU(N)$ theory with a hypermultiplet of mass $\mu$
in the bi-fundamental representation 
$({{\bf N}_1}, \overline {{\bf N}}_1)$ $ \oplus$ $(\overline
{{\bf N}} _1,{{\bf N}}_1)$
$$
\eqalign{
\F=&\F^{\rm class}+\F^{\rm pert}\cr
&+{1\over 2\pi i}\sum_{I=1,2}\bigg[
\sum_{i\in I}(-2\Lambda)^{N_1}S_i^I(x_i^I)
+{1\over 4}
(2\Lambda)^{2N_1}S_i^I(x_i^I)({\p\over\p x_i^I})^2S_i^I(x_i^I)\cr
&\qquad\qquad\qquad+(2\Lambda)^{2N_1}\sum_{i\not=j\atop i,j\in I}
{S_i^I(x_i^I)S_j^I(x_j^I)\over (x_i^I-x_j^I)^2}
\bigg].
\cr}
$$  

\medskip

\noindent
{\it The Case $N_2 < N_1$}

Substituting the limiting behavior of $q$ in (6.11), we find for the 
coefficients 
$$
\eqalign{
C_x = & 3(N_2 -2 N_1) + \pi i N_1 - N_2 \log 4 \cr
C_y = & 3(N_1 -2 N_2) + \pi i N_1 - N_1 \log 4 + 3 (N_1 - N_2)  \log { m^2
\over \Lambda ^2}
\cr}
$$
Clearly, $C_y$ diverges as $m \to \infty$, the coupling constant for the 
gauge group $SU(N_2)$ freezes out to zero and the fields $y_j, \ j=1,
\cdots , N_2$ are frozen to their constant expectation values. $C_x$ on the
other hand has a finite limit. Thus the remaining gauge group is the {\it
color group} $SU(N_1)$, while
$SU(N_2)$ effectively acts as a (approximate) {\it flavor group}.  Denoting 
$\mu - y _j \equiv m_j$, we readily recognize the curve (6.12) for this case
$$
A(x) - t B(x)  - 2^{N_2} \Lambda ^{2N_1 - N_2}  {1 \over t}  =0
\eqno (6.18)
$$
as the curve for an $\N=2$ supersymmetric theory with color group $SU(N_c)$,
$N_c=N_1$, and $N_f = N_2$ flavors in the fundamental representation of 
$SU(N_c)$. Notice that the models we have obtained this way are restricted
by $N_f < N_c$ !

\medskip

As in the previous case,
the effective prepotential for this model is also easily recovered from the
general expression for adjoint hypermultiplets in (5.23) up to 2 instanton
order by taking the limit $m\to \infty$, $\tau \to \infty$ with $\Lambda$
fixed 
and substituting into the general form (5.23). 
We obtain formulas similar to (6.17), with the expression
$\mu-x_j^2=m_j$ being interpreted this time however
as the mass $m_j$ of the hypermultiplet. 
The result clearly agrees
with the one obtained for this case in [7] for $N_c=N_1$ and $N_f=N_2$,
subject to the condition $N_f < N_c$, which is inherent to the construction
here.

\bigskip

\noindent
{\bf (c) Decoupling $SU(N) \to SU(N_1) \times \cdots \times SU(N_p)$}

\medskip

We now treat the general case of decoupling, in which $\tau$, $m$ and 
$k_i$ (or equivalently $a_i$) are all allowed to tend to infinity. Unless
there are special relations between the order parameters $k_i$, such limits
will either be free field theories or will be models containing several
mutually non-interacting systems. All such systems may be decomposed into
basic irreducible systems in which the $k_i$ form a {\it  linear chain}. 

\medskip

To define a  linear chain, we divide all the order parameters $k_i$ for
$i=1,\cdots , N=N_1 +\cdots + N_p$ into $p$ groups with $i=i_1=1, \cdots ,
N_1$ in the first group, $i = N_1 +i_2$ with $i_2 = 1 ,\cdots , N_2$ in the
second group and more generally $i = N_1 + N_2 + \cdots N_{I-1} + i_I$ with
$ i_I = 1, \cdots , N_I$ in the group indexed by $I=1,\cdots , p$. (We shall
assume that $N_I \geq 1$ for all $I=1, \cdots, p$ and use the notation $N_0
= N_{p+1} =0$ for convenience.) In analogy with (6.5), we decompose the
$k_i$ as follows
$$
k_{\{N_1 + \cdots + N_{I-1} + i_I\}} = v_I + x^I _{i_I}
\qquad \qquad
{I=1, \cdots , p \atop i_I = 1 ,\cdots , N_I}
\eqno (6.19)
$$
and fix the decomposition uniquely by requiring in addition that  $\sum
_{i_I=1} ^{N_I} x^I _{i_I} =0$, so that $\sum _I N_I v_I =0$. A linear chain
is such that as $m\to\infty$, the $x_{i_I}$ are fixed and the $v_I$ are
linked by the relations 
$$
v_I - v_{I+1} = m + \mu _I 
\qquad I = 1, \cdots , p-1
\eqno (6.20)
$$ 
where $\mu _I$ is also kept fixed. Notice that the definition of a linear 
chain  gives an ordering to the $N_I$. 

\medskip

\noindent
{\it Flow of Couplings, and Freezing out of the $U(1)$'s}

We shall now analyze the decoupling limit for an arbitrary linear chain, and
determine the behavior of $q$ as $m \to \infty$, as well as the conditions on the
integers $N_I$ for this theory to be irreducible.
We shall begin by obtaining the prepotential to classical and perturbative 
order for this arrangement of the order parameters. We define $\F ^{{\rm
class}} + \F ^{{\rm pert}} \equiv i \Phi /8 \pi$, and split up $\Phi$ into
a part with a manifestly finite limit
$$
\eqalign{
\Phi ^{{\rm lim}} 
= &\sum _{I=1}^p \sum_{i,j=1}^{N_I} 
   (x^I _i - x^I _j)^2 \log (x^I _i-x^I_j)^2/\Lambda^2
\cr
  & 
  -  \sum _{I=1} ^p \sum_{i=1}^{N_I} \sum _{j=1} ^{N_{I+1}} 
   (x^I_i - x^{I+1} _j + \mu _I )^2 \log (x^I _i-x^{I+1}_j  + \mu _I
)^2/\Lambda^2, 
\cr}
\eqno (6.21)
$$
a part which involves only quadratic dependence on $x ^I_i$ and only
quadratic and linear dependence on $\mu _I$
$$
\Phi ^{{\rm quad}} = C^0 _ {\mu _I} + \sum _{I=1} ^{p-1} C_{\mu _I} \mu _I
^2  - 4 \pi i \tau \sum _{I=1} ^{p-1} N_{I+1} (\mu _1 + \cdots +
\mu _I) ^2 +
\sum _{I=1} ^p    C_{x_I} \sum _{i=1 } ^{N_I} (x ^I_i) ^2 ,
\eqno (6.22)
$$
and a part which is independent of $x^I _{i_I}$ and $\mu _I$, and which may
be omitted. The coefficients $C_{x^I}$ and $C_{\mu _I}$ are given by
$$
\eqalign{
C_{x^I} = & C_{x^I} ^0 - 4 \pi i \tau + (N_{I+1} - 2 N_I + N_{I-1})  \log {
m^2 \over
\Lambda ^2}
\cr
C_{\mu _I} = &   N_I  N_{I+1} \log {m^2 \over \Lambda ^2}
\cr}
\eqno (6.23)
$$
The remaining coefficient $C^0 _{x ^I}$ depends only on the $N_I$ while 
the coefficient $C^0 _{\mu _I}$ depends upon $N_I$ as well as upon the
parameters $\mu _I$, but both are independent of $m$ and $\tau$. Neither
coefficients will be needed, and we shall not give them here.  

Now, it is immediately apparent that as soon as $\tau \to \infty$ and  $ m
\to\infty$,  we will have that all the coefficients $C_{\mu _I} \to \infty$
or more precisely $\Re (C_{\mu _I}) \to + \infty$. At the same time, the
real part of all coefficients in the third term in (6.22) also tend to
$+ \infty$. This means that all the couplings associated with the
$U(1)$ factors that would classically arise due to the breaking in (6.19) in
fact freeze out from the theory, and the true gauge group is reduced to
$SU(N_1)\times \cdots \times SU(N_p)$. The fields $\mu _I$ are frozen to 
their constant expectation values.

\medskip

Next, unless the coefficients $N_{I+1} - 2 N_I + N_{I-1}$ are all the same 
for $I=2, \cdots , p-1$, it will not be possible to choose a behavior for the
coupling $\tau$ such that the $p$ simple components $SU(N_I)$ remain mutually
interacting. In that case, the linear chain will break into two mutually
non-interacting smaller linear chains. We shall further justify this
assertion below when deriving the limiting curve. (The coefficients for
$I=1$ and
$I=p$ are not required to satisfy this condition since they correspond to
the groups at the end of the linear chain, and their freezing out will not
break the linear chain.) Thus, we shall assume that there exists an integer
$K$ such that
$$
N_{I+1} - 2 N_I + N_{I-1} = -K,
\qquad \qquad I= 2 , \cdots , p-1
\eqno (6.24)
$$
Clearly, for the couplings to have a finite limit, we need to have $K> 0$,
otherwise, each of the couplings associated with the groups $SU(N_I)$ will 
tend to zero as $\tau \to \infty$, and we would end up with a free theory.
Relation (6.24) requires a rather peculiar quadratic dependence of the $N_I$
on $I$, given for $I=2, \cdots , p-1$ by
$$
N_I = \2 K (I-1 ) (p-I) + { N_p - N_1 \over p-1} (I-1) + N_1,
\eqno (6.25)
$$
with two parameters $N_1$ and $N_p$, in addition to the number $p$. Notice
that (6.25) imposes constraints between these parameters, arising from
divisibility conditions : $2(N_p - N_1)$ must be divisible by $p-1$, and the
quotient must be even (odd) when $Kp$ is even (odd).

\medskip

\noindent
{\it The Limiting Curves}

Deriving the limiting curves for the $SU(N_1) \times \cdots \times SU(N_p)$
theories may be done in complete parallel to the case with just two factors,
carried out above. The polynomial $H(k)$, in terms of which the curve (4.5) 
is formulated, is calculated in the approximation of large $m$ first. We have
$$
H(x + v_1) = \prod _{I=1} ^p \prod _{i=1} ^{N_I} \big ( x + (I-1)m + M_I -
x^I _i \big )
\eqno (6.26)
$$
where we use the notation $M_I \equiv \mu _1 + \cdots + \mu _{I-1}$, with $
M_0=0$. The
$m\to
\infty$ limit is conveniently formulated in terms of the following $p$ 
polynomials
$$
A_I (x) \equiv \prod _{i=1} ^{N_I} (x + M_I - x_i ^I) 
\qquad 
I=1 , \cdots , p,
\eqno (6.27)
$$
and the following constants $c_n$, defined for given $N_I$ :
$$
\eqalign{
c_n \equiv & \prod _{I\not= 1+n} ^p (I -1-n) ^{N_I}
\qquad n= 0,1,\cdots, p-1 
\cr
c_n \equiv & \prod _{I=1} ^p (I -1-n) ^{N_I}
\qquad \quad n \not= 0,1,\cdots, p-1. 
\cr}
\eqno (6.28)
$$
The limits of $H(x+v_1 -nm)$ are then given by
$$
\eqalign{
H(x+v_1 -nm) = & c_n m^{N-N_{n+1}} A_{n+1}(x)
\qquad n=0,1, \cdots, p-1 \cr
H(x+v_1-nm) = & c_n m^N
\qquad \qquad \qquad \qquad 
\ n\not= 0,1,\cdots,p-1\cr}
\eqno (6.29)
$$
Just as for the case $p=2$, it is also necessary to redefine $w$ according to
$$
w \equiv t m^{N_2 - N_1}.
\eqno (6.30)
$$
Substituting this limiting behavior into the formula for the curve (4.5), 
we obtain
$$
\eqalign{
0= & 
  \sum _{n=0} ^{p-1} q ^{\2 n(n-1)} (-)^n t^n m^{N_1 - N_{n+1} +n(N_2-N_1)} 
c_n A_{n+1} (x)
\cr
 & + \sum _{n\not= 0,1,\cdots , p-1} ^{\infty} q^{\2 n(n-1)} (-)^n t^n 
m^{N_1+ n(N_2 - N_1)} c_n 
\cr}
\eqno (6.31)
$$
It is now clear, directly from equation (6.31) for the limiting curve, how
the quadratic dependence of $N_I$ on $I$ arises. In order to have a limit
where all gauge groups remain mutually interacting, we must retain the
dependence of the curve on all the polynomials $A_I(x)$ for all $I=1, \cdots
, p$. This can be achieved only if the combination $\Lambda$, defined by 
$$
q^{\2 n (n-1)} = \big ({ \Lambda  \over m} \big ) ^{ N_1 - N_{n+1} 
+ n (N_2 - N_1)},
\qquad n=0,1,\cdots , p-1,
\eqno (6.32)
$$ 
remains finite and can be held fixed. The equation is trivially satisfied 
for $n=0,1$, and for $n=2, \cdots , p-1$ reduces to (6.24) for unspecified
value of $K$. Henceforth, we shall assume that (6.24) is satisfied for some
$K>0$. The equation for the limiting curve then becomes
$$
0= 
  \sum _{n=0} ^{p-1}  (-)^n t^n \Lambda ^{\2 Kn(n-1)} c_n
        A_{n+1} (x)
 + S(t,m). 
\eqno (6.33)
$$
Here, the function $S(t,m)$ is defined by
$$
\eqalign{
S(t,m)
\equiv &\sum _{n\not= 0,1,\cdots, p-1} ^{\infty}  (-)^n t^n \Lambda^{\2 K
n(n-1)} m^{\nu _n } c_n  \cr
\nu _n \equiv
&  \ N_1 + n(N_2 - N_1) - \2 K n(n-1)
\cr}
\eqno (6.34)
$$
and $S(t,m)$ is independent on the order parameters $x^I _i$.

\medskip

The prepotential corresponding to (6.33) is again easily derived
by taking limits in (5.23). 
We may define its basic building blocks $A_i^I(x_i^I)$, $B^I(x_i^I)$,
$S_i^I(x_i^I)$ by exactly the same expressions as in (6.17).
Note however that for $1<I<p$, the corresponding expression $B^I(x_i^I)$
results now in
$$
B^I(x_i^I)
=\prod_{j\in I-1}(\mu_{I-1}+x_j^{I-1}-x_i^I)\prod_{j\in I+1}
(\mu_I+x_i^I-x_j^{I+1}).
$$
If we introduce the constants $\rho_I$ by
$$
\rho_I
=\prod_{j\in J\atop |J-I|\geq 2}(1-{1\over (I-J)^2}),
$$
we find the following limit for $i\in I$
$$
qS_i(x_i^I)\rightarrow
(-4)^{N_I}({\Lambda\over 2})^K\rho_IS_i^I(x_i^I),
$$
and hence the following expressions for the one- and two-instanton
corrections
$$
\eqalign{
\F^{(1)}&={1\over 2\pi i}({\Lambda\over 2})^K
\sum_I(-4)^{N_I}\rho_I\sum_{i\in I}S_i^I(x_i^I)\cr
\F^{(2)}&={1\over 2\pi i}({\Lambda\over 2})^{2K}
\sum_I4^{2N_I}\rho_I^2
\bigg[
\sum_{i\in I}S_i^I(x_i^I)({\p\over\p x_i^I})^2S_i^I(x_i^I)
+
\sum_{i,j\in I\atop i\not=j}
{S_i^I(x_i^I)S_j^I(x_j^I)\over (x_i^I-x_j^I)^2}
\bigg].
\cr}
$$
Again, as in Section VI.(b), some of the parameters $x_i^I$ may
freeze to a constant expectation value, and the resulting $SU(N_I)$
should be viewed rather as a flavor group. We shall see below
that this can take place only for $SU(N_p)$ and $SU(N_1)$.

\medskip

The dynamics of the theory may be divided up into three categories, which 
are specified by the values of the parameters $N_1$ and $N_p$. Actually,
the combinations that enter more naturally (and are equivalent to $N_1$ and
$N_p$) are given by
$$
K^+ \equiv  \ 2N_p - N_{p-1}
\qquad \qquad
K^- \equiv  \ 2N_1 - N_2,
\eqno (6.35)
$$
and (6.25) may equivalently be expressed in terms of these for $I=2, \cdots
, p-1$
$$
N_I = \2 K I (p+1 -I)  + { 1 \over p+1} \big ( I K ^+ + (p+1 -I)  K^- -(p+1)
K
\big ).
\eqno (6.36)
$$
We now have the condition that $2(K^+ - K^-)$ be divisible by $p+1$ and
that the resulting quotient be even (odd) if $Kp$ is even (odd).

\medskip

The function $S(t,m)$ of (6.34) will have a convergent limit if and only if
$\nu _n$ of (6.34) satisfies  
$$
\nu _n \leq 0 
\qquad \qquad 
{\rm for \ all \ } n\in {\bf Z}, \ n \not= 0,1, \cdots, p-1.
\eqno (6.37)
$$
Since $\nu _n$ is a downward parabola as a function of $n$, with $\nu _0 =
N_1 >0$, it is necessary and sufficient for (6.37) to hold that 
$$
\nu _{-1} =  K^- - K \leq 0
\qquad \qquad  
\nu _{+p} =  K^+ - K \leq 0. 
\eqno (6.38)
$$
Assuming that $S(t,m)$ indeed converges to a finite limit, we are left with
three inequivalent cases, which we now describe in turn.

\medskip

\noindent
{\it (1) The Case $K^+ = K^- = K$ : Gauge Group $SU(N_1) \times \cdots \times
SU(N_p)$ with  Hypermultiplets in bi-Fundamental Representations
}

The couplings of all $SU(N_I)$ have a finite limit as $m,\ \tau \to\infty$,
and the full gauge group  $SU(N_1) \times \cdots \times SU(N_p)$ remains. The
formula for the $N_I$ simplifies and we have $N_I = \2 K I (p+1 -I)$ for all
$I=1, \cdots , p$. The $p-1$ hypermultiplets that remain after decoupling
are in bi-fundamental representations, given by 
$$
\sum _{I=1} ^{p-1} \big \{({\bf N}_I, \overline {{\bf N}}_{I+1}) \oplus 
     (\overline {{\bf N}}_I, {{\bf N}}_{I+1}) \big \}.
\eqno (6.39)
$$
The limiting curve results from (6.33) with the finite limit of $S(t,m)$,
which comes from the $n=-1$ and $n=p$ contributions only :
$$
0= 
  \sum _{n=0} ^{p-1}  (-)^n t^n \Lambda ^{\2 K n(n-1)} c_n
        A_{n+1} (x)
 - {1 \over t} \Lambda ^K c_{-1} + (-) ^p t^p \Lambda ^{\2 Kp(p-1)} c_p. 
\eqno (6.40)
$$
The cases $N_1=N_2$, treated above are of this type with $p=2$.

\medskip

\noindent
{\it (2) The Case $K^+< K ,\ ~K^- =K$ : Gauge Group $SU(N_1) \times \cdots
\times SU(N_{p-1})$ with Hypermultiplets in Fundamental
\&  bi-Fundamental Representations  }

The cases $K^+< K ,\ K^- = K$ and $K^-< K ,\ K^+ = K$ are clearly
equivalent, and we shall limit to dealing with the first. In that case, the
limits of the couplings of $SU(N_1) \times \cdots \times SU(N_{p-1})$ are
finite, while the coupling of $SU(N_p)$ converges to zero, so that the
dynamics of this group gets frozen out. The group $SU(N_p)$ effectively
becomes a flavor group and we have the following hypermultiplet contents
: $p-2$ hypermultiplets in bi-fundamental representations, together with
$N_p$ hypermultiplets in the fundamental representations of each of the
simple factors of the gauge group. In total, we have
$$
\sum _{I=1} ^{p-2}  \bigl \{ ({\bf N}_I, \overline {{\bf N}}_{I+1})
\oplus 
     (\overline {{\bf N}}_I, {\bf N}_{I+1}) \bigr \}  \oplus \sum _{I=1}
^{p-1} N_p \big \{{\bf N}_I \oplus \overline{{\bf N}} _I\big \}
\eqno (6.41)
$$
The curve is easily obtained from (6.33) again and the limit now only
contains a single term in the function $S(t,m)$, namely for $n=-1$. We find
$$
0= 
  \sum _{n=0} ^{p-1}  (-)^n t^n \Lambda ^{\2 K n(n-1)} c_n
        A_{n+1} (x)
 - {1 \over t} \Lambda ^K c_{-1} 
\eqno (6.42)
$$
The prepotential to 2 instanton order is again obtained by starting from the
general formula (5.23) and taking the above limit $m \to \infty$ and $\tau
\to \infty$.

\medskip

\noindent
{\it (3) The Case $K^- ,K^+<K$ : Gauge Group $SU(N_2) \times \cdots \times
SU(N_{p-1})$ with Hypermultiplets in  Fundamental \& bi-Fundamental
Representations } 

In this case, only the couplings of the gauge group  $SU(N_2) \times \cdots
\times SU(N_{p-1})$ converge to a finite value, while those of
$SU(N_1)\times SU(N_p)$ converge to zero. Thus, $SU(N_1)\times
SU(N_p)$ becomes an approximate flavor group. There are now $p-3$
hypermultiplets in bi-fundamental representations (for $p=2$, there are no
such hypermultiplets), and $N_1 + N_p$ hypermultiplets in each of the
fundamental representations of the simple factors of the gauge group. In
total, we have
 $$
\sum _{I=2} ^{p-2}  \bigl \{ 
({\bf N}_I, \overline {{\bf N}}_{I+1})
\oplus 
     (\overline {{\bf N}}_I, {\bf N}_{I+1}) \bigr \}  
\oplus \sum _{I=2}^{p-1}
( N_1 + N_p) \big \{{\bf N}_I \oplus \overline{{\bf
N}} _I\big \}
\eqno (6.43)
$$
The curve for this theory is
$$
0= 
  \sum _{n=0} ^{p-1}  (-)^n t^n \Lambda ^{\2 K n(n-1)} c_n
        A_{n+1} (x)
\eqno (6.44)
$$

\bigskip

\noindent
{\bf (d) The Special Case $p=3$}

\medskip

It is very instructive to see how the general discussion of (c) can be
applied to the special case $p=3$. For solutions to exist, it is necessary
and sufficient that $K>0$ and that the following general formal solutions be
positive integers
$$
\eqalign{
N_1 = & \2 K + \4 (K^+ + 3 K^-) \cr
N_2 = & K + \2 (K^+ + K^-) \cr
N_3 = & \2 K + \4 (3 K^+ + K^-) \cr
N= & 2K + {3 \over 2} (K^+ + K^-). \cr}
\eqno (6.45)
$$
A particularly interesting limit is where $K^+, K^- <K$, so that the gauge
groups $SU(N_1)$ and $SU(N_3)$ freeze out. Let us call the remaining gauge
group color and set $N_c=N_2$ to be the number of colors. According to our
general discussion, the freezing out of the gauge groups produces $N_f=N_1
+ N_3 = K + K^+ + K^-$ flavors of fundamental representations of $SU(N_c)$.
The quantity $2N_c - N_f = K$ had to be positive from general
considerations and we now recognize this conditions as the criterion for
asymptotic freedom ! Let's check that we obtain the curve (3.1). The curve
equation that emerges from (6.44) is given by
$$
c_0 A_1(x) - t c_1 A_2 (x) + t^2 c_2 \Lambda ^{2 N_c - N_f} A_3(x) =0
\eqno (6.46)
$$
This equation would appear to be different from (3.1). However, we shall
now make the following change of variables
$$
t=   { A_1(x) \over \tilde y} { c_0 \over c_1} 
$$
and define $A(x) = A_2(x)$, $B(x) = A_1(x) A_3(x)$, which puts the equation
in a form 
$$
\tilde y ^2 - \tilde y A(x) + {c _0  c_2 \over c_1 ^2} \Lambda ^{2N_c - N_f}
B(x) =0.
\eqno (6.47)
$$
This is equivalent to (3.1), upon setting $y=2\tilde y -A(x)$ and
redefining $\Lambda$ by a multiplicative constant. The only restrictions on
this result were $K^+ ,\ K^- <K$, so that the above construction is limited
to $2N_c - N_f \geq 2$.

\bigskip
\bigskip

\centerline{\bf ACKNOWLEDGEMENTS}

\bigskip

The authors would like to thank especially I.M. Krichever
for collaboration on many closely related topics,
and for sharing with them his insights on Calogero-Moser systems. 
They have also benefited from conversations with Elena Caceras, Ron Donagi, 
and Edward Witten.
One of them (E.D.) would like to thank Edward Witten
for inviting him to spend the spring 1997 semester at
the Institute for Advanced Study, where part of this work was carried out.
He would like to acknowledge the generous support from the
Harmon Duncombe Foundation during this period. The authors would 
also like to thank
the Aspen Center for Physics, and the Mathematical Sciences Research
Institute of Berkeley, where the final part of this work was done.

\vfill\break

\centerline{\bf REFERENCES}

\bigskip

\item{[1]} Seiberg, N. and E. Witten, ``Monopoles, duality,
and chiral symmetry breaking in N=2 supersymmetric QCD",
Nucl. Phys. {\bf B431} (1994) 494, hep-th/9410167.

\item{[2]} Donagi, R. and E. Witten, ``Supersymmetric Yang-Mills
theory and integrable systems", Nucl. Phys. {\bf B460} (1996) 299-334,
hep-th/9510101.

\item{[3]} Krichever, I.M., ``Elliptic solutions of the
Kadomtsev-Petviashvili equation and integrable systems
of particles", Funct. Anal. Appl. {\bf 14} (1980) 282-290.

\item{[4]} Martinec, E., ``Integrable structures in supersymmetric
gauge and string theory", hep-th/9510204.

\item{[5]} Krichever, I.M. and D.H. Phong, ``On the integrable geometry
of soliton equations and N=2 supersymmetric gauge theories",
J. of Differential Geometry {\bf 45} (1997) 349-389.

\item{[6]} Krichever, I.M. and D.H. Phong, ``Symplectic forms in the
theory of solitons", hep-th/9708170,
submitted to Surveys in Differential Geometry, Vol. III.

\item{[7]} D'Hoker, E., I.M. Krichever, and D.H. Phong,
``The effective prepotential for N=2 supersymmetric
SU($N_c$) gauge theories", Nucl. Phys. {\bf B 489} (1997) 179-210,
hep-th/9609041; ``The effective prepotential
for N=2 supersymmetric SO($N_c$) and Sp($N_c$) gauge theories",
Nucl. Phys. {\bf B 489} (1997) 211-222, hep-th/9609145.

\item{[8]} D'Hoker, E. and D.H. Phong, ``Strong coupling expansions in
SU(N) Seiberg-Witten theory", Phys. Lett. {\bf B 397} (1997) 94-103.

\item{[9]} Klemm, A., W. Lerche, and S. Theisen, ``Non-perturbative
actions of N=2 supersymmetric gauge theories",
Int. J. Mod. Phys. {\bf  A11} (1996) 1929-1974, hep-th/9505150. 

\item{[10]} Lerche, W., ``Introduction to Seiberg-Witten theory
and its stringy origin", Proceedings
of the {\it Spring School and Workshop in String Theory},
ICTP, Trieste (1996), hep-th/9611190.

\item{[11]} D'Hoker, E., I.M. Krichever, and D.H. Phong,
``The renormalization group equation for N=2 supersymmetric
gauge theories", Nucl. Phys. {\bf B 494} (1997), 89-104, hep-th/9610156.

\item{[12]} Witten, E., ``Solutions of four-dimensional
field theories via M-theory", IASSNS-HEP-97-19, March 1997.

\item{[13]} Hitchin, N., ``Stable bundles and integrable systems",
Duke Math. J. {\bf 54} (1987) 91-114.

\item{[14]} Krichever, I.M., O. Babelon, E. Billey, and M. Talon,
``Spin generalization of the Calogero-Moser system and the
matrix KP equation", Amer. Math. Soc. Trans. {\bf 170} (1995) 83-119.

\item{[15]} Krichever, I.M., ``The $\tau$-function of the universal
Whitham hierarchy, matrix models, and topological
field theories", Comm. Pure Appl. Math. {\bf 47} (1994) 437-475.

\item{[16]} Sonnenschein, J., S. Theisen, and S. Yankielowicz,
Phys. Lett. {\bf B367} (1996) 145-150, hep-th/9510129.

\item{[17]} Eguchi, T. and S.K. Yang, ``Prepotentials
of N=2 supersymmetric gauge theories and soliton
equations", hep-th/9510183.

\item{[18]} Matone, M., Phys. Lett. {\bf  B357} (1996) 342.

\item{[19]} Gorski, A., I.M. Krichever, A. Marshakov, A. Mironov, A.
Morozov, Phys. Lett. {\bf B355} (1995) 466, hep-th/9505035; \hfill\break
Nakatsu, T. and K. Takasaki, Mod. Phys. Lett. {\bf A 11}
(1996) 157-168, hep-th/95\-09\-162; \hfill\break
Itoyama, H. and A. Morozov, ``Prepotential
and the Seiberg-Witten theory", hep-th/9512161; ``Integrability
and Seiberg-Witten theory", hep-th/9601168;\hfill\break
Marshakov, A. ``Non-perturbative
quantum theories and integrable equations",
hep-th/9610242; \hfill\break
Ahn, C. and S.Nam, hep-th/9603028. 

\item{[20]} Friedan, D., ``Introduction to Polyakov's string theory",
in {\it Recent Advances in Field Theory and
Statistical Mechanics}, Les Houches (1982), edited by
J.B. Zuber and R. Stora, North Holland, Amsterdam, 839.

\item{[21]} D'Hoker, E. and D.H. Phong, ``The geometry of string
perturbation theory", Rev. Mod. Physics {\bf 60} (1988) 917-1065.

\item{[22]} D'Hoker, E. and D.H. Phong, ``Conformal scalar fields
and chiral splitting on super Riemann surfaces",
Comm. Math. Phys. {\bf 125} (1989) 469-513.

\item{[23]} Perelomov, A.M., ``Integrable Systems of Classical Mechanics
and Lie Algebras", Vol. I, Birkhauser Verlag (1990); and references therein.

\item{[24]} Martinec, E. and Warner, N., ``Integrable systems and
supersymmetric gauge theories", Nucl. Phys. {\bf B459} (1996) 97-112,
hep-th/9509161.

\item{[25]} Katz, S., P. Mayr, and C. Vafa, ``Mirror symmetry
and exact solutions of 4D N=2 gauge theories",
hep-th/9706110.

\end